\documentstyle[psfig,epsf,12pt]{article}
\begin{document}
\title{Light-cone Hamiltonian flow for positronium.\\
       The numerical solutions}
\author{Elena L. Gubankova$^{1}$\thanks{Present address: Department of
Physics, North Carolina State University, Raleigh, 27695-8202 NC USA}
  and G\'abor Papp$^{1,2}$\thanks{on leave from 
 Inst. for Theor. Physics, E\"otv\"os University, P\'azm\'any P.s. 1/A,
 Budapest, H-1117} \\
$^1$ Institut f\"ur Theoretische Physik der Universit\"at Heidelberg, \\ 
   D-69120 Heidelberg \\
$^2$ CNR, Department of Physics, Kent State University, \\
     44242 OH USA 
}
\date{15 January, 1999}
\maketitle
\begin{abstract}
The effective Hamiltonian, as obtained from applying 
the Hamiltonian flow equations to front form QED,
are solved numerically for positronium.
Both the exchange and the annihilation channels are included.
The impact of different similarity functions is explicitly studied.
Perfect numerical agreement with other methods is found.
\end{abstract}
\vfill

\newpage\tableofcontents\newpage

\section{Introduction}
\label{sec:1}

Light-front frame is believed to be a useful tool for solving bound state 
problems in QCD. Generally, bound state calculations in the field theory
have two sources of complexity - they are relativistic and of many-body type.
The method of flow equations copes with both of them, at least to the definite
order in perturbation theory. One transforms the Hamiltonian to eliminate
interactions changing particle number, reducing thus the bound state
problem to a few-body problem. Simultaneously utraviolet divergencies
occur, originating from the high-energy region. To complete
renormalization one uses either coupling coherence or
fixes counterterms to provide finite values for physical observables
and to retain symmetries violated by the procedure~\cite{Perry}.
Both type of flow equations of renormalization type and for the new
(particle number conserving) interactions appear together.
This program can be fulfilled in perturbation theory expansion. 
As a result, the bound state problem
is approximated by a set of renormalized, effective interactions
that do not change particle number.

The sensitive tool in the light-front frame to check how accurately
one desribes bound states by these effective interactions is to measure
the violation of rotational symmetry. This symmetry is linked to a
dynamical operator on the light-front, since rotations are dynamical,
i.e. depend on the interaction. The symmetry may be spoiled in two
steps: first, regularization and renormalization;
and second, reduction to the effective few-body interactions with
particle number conservation.
The nonperturbative renormalization flow is of crucial importance
for QCD~\cite{BrPe,Perry,pau96}, but one can disregard
this point in QED bound state calculations, that disentangles the two problems
mentioned above.
To the leading order the results for QED are obtained in~\cite{JoPeGl},
where positronium system is described approximately by the effective 
electron-positron interaction. In the nonrelativistic limit
the results for positronium spectrum agree with the results of covariant
calculations. 

There are at least two other alternative approaches to solve for bound states
in the light-front dynamics, the scheme of similarity renormalization of Glazek
and Wilson~\cite{GlWi} and the method of iterated resolvents of Pauli~\cite{pau96}.
In both schemes calculations of the effective electron-positron
interaction are performed and the question of rotational invariance for
positronium spectrum is investigated.

Calculations done so far in the similarity renormalization scheme use
the nonrelativistic limit to find corresponding eigenvalues in the bound
state perturbation theory~\cite{BrPe2}.  Analytical calculations are
performed there for ground state: ground triplet levels are degenerate,
indicating that rotational symmetry is restored~\cite{BrPe2}. Performing
similarity renormalization one eliminates high-energy modes and absorbs
relativistic effects into an effective band-diagonal Hamiltonian, which
describes bound state creation at nonrelativistic energy scales.  It is
a well working scheme for such systems as positronium~\cite{Perry};
therefore nonrelativistic approximations done in this approach to
extract eigenvalues from effective Hamiltonian are quite natural
there. In general, it is not always the case.  In fact rotational
symmetry becomes kinematic one, like light-front boost in the
nonrelativistic limit, i.e. total momentum and its projection can be
considered approximately as quantum numbers, that makes simpler to trace
rotational invariance in these calculations.

In the method of iterated resolvents an effective electron-positron
potential is obtained and exact numerical solution of positronium bound
state equation with the given potential is done~\cite{TrPa,TrPa2}.  
Degenerate multiplets for ground as well as for exited states are
obtained~\cite{TrPa}. \footnote{
Numerical solution of positronium bound state problem in the light-front frame
can be found also in ~\cite{kpw92,KaPi}.  }
It is convenient to perform relativistic calculations
in the light-front frame, which effectively has nonrelativistic
kinematics.

In the present work we perform relativistic few-body calculations for
positronium spectrum numerically in the spirit of the work Trittmann
et.al.~\cite{TrPa} (and using numerical code ~\cite{TrPa2}), based on the
effective electron-positron Hamiltonian obtained by the flow
equations~\cite{previous}.  Effective interaction was derived there for
different cutoff functions.  The requirement of block-diagonalization of
the Hamiltonian determines the generator only up to a unitary
transformation of the blocks; this explains why the effective
interaction may depend on the cutoff function.  The question we
investigate is to what extent rotational invariance is violated on the
level of positronium spectrum and how does it depend on the choice of
the cutoff function.  We are not able to trace rotational invariance
during the calculations, since it is dynamical operator.  This is an
excellent test for the method of flow equations itself and the control of
approximations done during the calculations.

\section{Formulation of the problem}
\label{sec:2}

We address to solve a light-front Hamiltonian bound state equation
\begin{equation}
    H \vert\psi\rangle =E \vert\psi\rangle
\label{eq:i1}\end{equation}
for positronium. Using flow equations we transform the QED Hamiltonian $H$  
to a block-diagonal effective Hamiltonian, which reduces positronium problem 
to a bound state problem in the electron-positron sector. 
The effective Hamiltonian for an electron and a positron is
\begin{equation} 
    H_{\rm eff}=H_0+U_{\rm eff}
\label{eq:i2}\end{equation}
where $H_0$ is the kinetic energy, and $U_{\rm eff}$ includes effective 
interactions generated by the flow equations in the second order in
coupling constant.
The integral bound state equation is written
\begin{eqnarray}
    E \langle p_1,p_2; \lambda_1,\lambda_2 \vert \psi \rangle &=&
    (E_{p_1}\!+\!E_{p_2}) 
	\langle p_1,p_2; \lambda_1,\lambda_2 \vert \psi \rangle 
\nonumber\\
    &+& \sum_{\lambda_1^\prime,\lambda_2^\prime}
    \!\int\! d^3p^\prime_1 d^3p^\prime_2 \langle p_1,p_2; \lambda_1,\lambda_2  
    \vert U_{\rm eff}\vert p^\prime_1,p^\prime_2; 
		\lambda_1^\prime,\lambda_2^\prime \rangle
    \langle p^\prime_1,p^\prime_2;\lambda_1^\prime,
		\lambda_2^\prime \vert \psi \rangle
\nonumber\\ 
\label{eq:i3}\end{eqnarray}
where the effective Hamiltonian pickes out from the positronium wave function
$\vert \psi \rangle$ the lowest $e\bar{e}$-component 
$\langle p_1,p_2;\lambda_1,\lambda_2 \vert \psi \rangle$
with $p_i,\lambda_i$ being the light-front three-momenta
and helicities, respectively, carried by an electron ($i=1$), and a
positron ($i=2$). The primed quantities refer to the initial state,
the unprimed ones to the final state. The effective interaction
$ \langle p_1,p_2;\lambda_1,\lambda_2 \vert U_{\rm eff}
  \vert p^\prime_1,p^\prime_2;\lambda_1^\prime,\lambda_2^\prime \rangle = 
  U_{\rm eff} \delta(p_1+p_2-p'_1-p'_2) $ will be specified below.
In order to deduce a Lorentz invariant energy we consider the bound
state equation written for operator $P^-P^+$, corresponding to the
invariant mass-squared $M^2$ on the light-front, rather than for the 
light-front Hamiltonian operator $H=P^-$.
The light-front integral equation,
\\
\begin{eqnarray} 
	M^2\ \langle x,\vec \kappa_{\!\perp}; \lambda_1,
    \lambda_2  \vert \psi \rangle &=&
    {m^2 \!+\! \vec \kappa_{\!\perp}^2 \over x(1\!-\!x)} 
    \ \langle x,\vec \kappa_{\!\perp}; \lambda_1,
    \lambda_2  \vert \psi \rangle 
\nonumber\\
    &+& \sum _{ \lambda_1^\prime,\lambda_2^\prime}
    \int\limits_D\!dx^\prime d^2 \vec \kappa_{\!\perp}^\prime\,
    \,\langle x,\vec \kappa_{\!\perp}; \lambda_1, \lambda_2
    \vert V_{\rm eff}
    \vert x^\prime,\vec \kappa_{\!\perp}^\prime; 
    \lambda_1^\prime, \lambda_2^\prime\rangle
    \ \langle x^\prime,\vec \kappa_{\!\perp}^\prime; 
    \lambda_1^\prime,\lambda_2^\prime  
    \vert \psi \rangle
\nonumber\\
\label{eq:r96}
\end {eqnarray}
is independent of the total momentum $P^+$ and $\vec P_\perp$. We
introduced $V_{\rm eff}=P^{+2}U_{\rm eff}$. 
In that equation only intrinsic transversal momenta
$\vec \kappa_{\!\perp}$ and longitudinal momentum fractions 
$x = p_1^+/P^+$ appear ($p_1^\mu=(xP^+,x\vec P_\perp + \vec
\kappa_{\!\perp},p_1^-)$).
Its spectrum is thus manifestly independent of the kinematical state
of the bound system, particularly of $P^+$ and $\vec{P}_{\perp}$,
which reflects on the boost invariance peculiar to the light-front
form~\cite{bpp97}. 
The integration domain $D$ is restricted by the covariant cutoff
condition of Brodsky and Lepage~\cite{LeBr},
\begin{eqnarray}
 \frac{m^2+\vec{\kappa}_{\perp}^{2}}{x(1-x)}\leq \Lambda^2+4m^2
\,,\label{eq:domain}\end{eqnarray} 
which allows for states having a kinetic energy below the bare 
cutoff~$\Lambda$.
The effective interaction between electron and positron,
being a kernel in the integral equation~(\ref{eq:r96}),
is generated by the flow equations~\cite{previous}
\begin{eqnarray}
   V_{\rm eff} = &-& 
   \frac{\alpha}{4\pi^2} \langle \gamma^\mu\gamma^\nu\rangle_{ex}
   \left[g_{\mu\nu}\,
   \left(\frac{\Theta_{e \bar e}} {Q_{e}^2} +
         \frac{\Theta_{\bar e e}} {Q_{\bar e}^2}\right) + 
   \eta_\mu\eta_\nu\,\frac {\delta Q^2}{{q^+}^2} 
   \left(\frac{\Theta_{e \bar e}} {Q_{e}^2} - 
         \frac{\Theta_{\bar e e}} {Q_{\bar e}^2}\right)
   \right]
\nonumber\\
  &-& \frac{\alpha}{4\pi^2} \langle \gamma^\mu\gamma^\nu\rangle_{an}
  \left[
   g_{\mu\nu}
   \left(\frac{\Theta_{ab}} {M_a^2} +
         \frac{\Theta_{ba}} {M_b^2} \right) - 
   \eta_\mu\eta_\nu\,\frac{\delta M^2} {{p^+}^2}
   \left(\frac{\Theta_{ab}} {M_a^2} - 
         \frac{\Theta_{ba}} {M_b^2} \right)  \right]
\,.\label{eq:56}\end{eqnarray}
with the generator of unitary transformation
\begin{eqnarray}
 \eta(l) &=& -\frac{1}{D}\left(\frac{d{\rm ln}\,f(D;l)}{dl}\right)g(l)
\,.\label{eq:gen}\end{eqnarray}
where $g(l)$ is the coupling constant 
as a function of flow parameter $l$, and $f(D;l)$ is the cutoff function
specified below.
In Eq.~(\ref{eq:56}) subscript $ex$ refers to the exchange part, 
and $an$ to the annihilation part. 
The null vector $\eta^{\mu}$ has components
$(\eta^+,\vec\eta_\perp, \eta^-)=(0,\vec{0},2)$ and is specific 
to the light-front calculations. 
The light-front metric tensor is denoted by $g_{\mu\nu}$.
The current-current tensors in the two channels are
\begin{eqnarray}
   \langle \gamma^\mu\gamma^\nu\rangle_{ex} &=& 
   \frac{(\overline u(p_1,\lambda_1) \gamma^\mu u(p'_1,\lambda'_1))\,
   (\overline v(p'_2,\lambda'_2) \gamma^\nu v(p_2,\lambda_2))}
   {\sqrt{x x' (1-x) (1-x')} }
\,,\nonumber\\
   \langle \gamma^\mu\gamma^\nu\rangle_{an} &=&
   \frac{(\overline u(p_1,\lambda_1)\gamma^\mu  v(p_2,\lambda_2))\,
   (\overline v(p'_2,\lambda'_2)\gamma^\nu u(p'_1,\lambda'_1))}
   {\sqrt{x x' (1-x) (1-x')} }
\,,\label{eq:r94}\end{eqnarray}
where the fermion momenta were defined after Eq.~(\ref{eq:i3}). 
The remaining definitions are as follows.
The energy differences along the electron 
and the positron line, 
\begin{eqnarray}
   D_{e} &=& {p'_1}^- - p_1^{-} - (p'_1-p_1)^-
\,,\nonumber\\
   D_{\bar e} &=& p_2^{-} - {p'_2}^- - (p_2-p'_2)^-
\,.\label{eq:40a} \end{eqnarray}
respectively, have a simple relation 
to the (Feynman-) 4-momentum transfers along the two lines 
\begin{eqnarray}
   Q_{e}^2 &=& -(p'_1-p_1)^2 = -q^+D_{e} 
\,,\nonumber\\
   Q_{\bar e}^2 &=& -(p_2-p'_2)^2 = -q^+D_{\bar e}
\,.\end{eqnarray}
Since the Feynman-momentum transfer $Q$ is more physical
quantity than the energy difference, we will use the former
as far as possible.
In fact, in our formulae we make use of the
{\em mean-square momentum transfer} and the {\em mean-square difference},
\begin{eqnarray}
        Q^2 &=& {1\over 2}(Q_{e}^2+Q_{\bar e}^2)
        = -{q^+\over 2}(D_{e}+D_{\bar e})
\,,\nonumber\\  
 \delta Q^2 &=& {1\over 2}(Q_{e}^2-Q_{\bar e}^2)
        = -{q^+\over 2}(D_{e}-D_{\bar e})
\,,\label{eq:r15}\end{eqnarray}
respectively. The dependence of the effective interaction Eq.~(\ref{eq:56})
on the cutoff function $f(D;l)$ is carried by the factor
\begin{equation}
   \Theta(D_{e},D_{\bar e}) = - \int_{0}^\infty\!dl'\,
   \frac{d f (D_{e};l')}{dl'} f (D_{\bar e};l')
   \equiv \Theta_{e\bar e}
\,,\label{eq:r16}\end{equation}
which is asymmetric in the arguments but which 
satisfies
\begin{equation}
   \Theta(D_{e},D_{\bar e})+\Theta(D_{\bar e},D_{e}) =
   \Theta_{e\bar e} + \Theta_{\bar e e} = 1
\,.\label{eq:ii18a}\end{equation}
The latter combination obeys
\begin{eqnarray}
   \frac{\Theta_{e \bar{e}}} {Q_{e}^2}+
   \frac{\Theta_{\bar{e} e}} {Q_{\bar{e}}^2} &=& 
   \frac {Q^2} {Q_{e}^2 Q_{\bar{e}}^2} 
   \left( 1 - \frac{\delta Q^2}{Q^2} \left(\Theta_{e\bar{e}} - 
   \Theta_{\bar{e}e} \right) \right) 
\nonumber\\
   \frac{\Theta_{e\bar{e}}} {Q_{e}^2}-
   \frac{\Theta_{\bar{e}e}}{Q_{\bar{e}}^2} &=& - 
   \frac {\delta Q^2} {Q_{e}^2 Q_{\bar{e}}^2} 
   \left( 1 - \frac{ Q^2}{\delta Q^2} \left(\Theta_{e\bar{e}} - 
   \Theta_{\bar{e}e} \right) \right) 
\,.\label{eq:ii18b}\end{eqnarray}
that we use further.

For the annihilation term we define the energy differences as
\begin{eqnarray}
   D_a &=& {p'_1}^- + {p'_2}^- - (p'_1+p'_2)^-
\nonumber\\
   D_b &=& p_1^{-} + p_2^{-} - (p_1+p_2)^-
\,.\label{eq:40b} \end{eqnarray}
They are related to the 4-momentum $p^\mu$ of the photon 
and to the free invariant 
mass-squares of the initial and final states 
\begin{eqnarray}
   M_a^2 &=& (p'_1+p'_2)^2 = p^+ D_a 
\,,\nonumber\\
   M_b^2 &=& (p _1+p _2)^2 = p^+ D_b
\,,\end{eqnarray}
as well as to their mean and difference
\begin{eqnarray}
        M^2 &=& {1\over 2}(M_a^2 + M_b^2)= {p^+\over 2}(D_a+D_b)
\,,\nonumber\\  
 \delta M^2 &=& {1\over 2}(M_a^2 - M_b^2)= {p^+\over 2}(D_a-D_b)
\,,\label{eq:r16a}\end{eqnarray}
respectively. 
 
Effective interaction~(\ref{eq:56}) includes two different
Lorentz structures: $g_{\mu\nu}$ part insures the Bohr spectrum
and is responsible for the spin splittings; 
$\eta_{\mu}\eta_{\nu}$ term is diagonal in spin space
and vanishes for real processes, i.e. on mass shell with $\delta Q^2=0$,
making the effective interaction to
coincide with the Tamm-Dancoff approximation ~\cite{previous}.
The explicit $x$-dependence in the denominator of Eq.~(\ref{eq:r94})
looks like the only remnant of the light-front formulation;
all other quantities are Lorentz scalars.
One can absorb this dependence by redefining the wave function 
in the integral equation Eq.~(\ref{eq:r96}). 
We introduce instead
of Jacobi momentum $(x,\vec{\kappa}_{\perp})$ the three momentum
in the center of mass frame $\vec{p}=(p_z,\vec{\kappa}_{\perp})$
as follows
\begin{eqnarray}
   x = \frac{1}{2}\left(1+\frac{p_z}{\sqrt{{\vec p}^{\, 2}+m^2}} \right)
\,,\label{eq:eq3}\end{eqnarray}
where the Jacobian of this transformation $dx/dp_z$ is
\begin{eqnarray}
   J = \frac{1}{2}\frac{\vec{\kappa}_{\perp}^2+m^2}
                       {({\vec p}^{\, 2}+m^2)^{3/2}}
     = 2\frac{x(1-x)}{E}
\,,\label{eq:eq4}\end{eqnarray}
and it holds in this frame  
\begin{eqnarray}
   x(1-x) &=& \frac{1}{4}\frac{\vec{\kappa}_{\perp}^2+m^2}{{\vec p}^{\, 2}+m^2}
\,,\nonumber\\
   E &=& \sqrt{{\vec p}^{\, 2}+m^2}
\,.\end{eqnarray}
The connection between `old' and `new' wave functions and the
interaction matrix elements are
\begin{eqnarray} 
   \langle x,\vec{\kappa}_{\perp} \vert \psi \rangle &=&
   \frac{\langle \vec{p} \vert \psi^\prime \rangle}{\sqrt{x(1-x)}}
\,,\nonumber\\
   \langle x,\vec \kappa_{\!\perp} \vert V_{\rm eff}
    \vert x^\prime,\vec \kappa_{\!\perp}^\prime \rangle &=&
    \frac{ \langle \vec {p} \vert V_{\rm eff}^\prime 
    \vert {\vec p}^{\: \prime} \rangle}
    {\sqrt{x(1-x)x'(1-x')}}
\,.\end{eqnarray}
Integral equation~(\ref{eq:r96}),
\begin{eqnarray} 
    M^2\ \langle \vec{p}; \lambda_1,
    \lambda_2  \vert \psi^\prime \rangle &=&
    4 ({\vec p}^{\, 2}+m^2)
    \ \langle \vec p; \lambda_1,
    \lambda_2  \vert \psi^\prime \rangle 
\nonumber\\ 
&+& \sum _{ \lambda_1^\prime,\lambda_2^\prime}
    \!\int_D\! \frac{d^3 {\vec p}^{\: \prime}}{2E}\,
    \,\langle \vec p; \lambda_1, \lambda_2
    \vert V_{\rm eff}^\prime
    \vert {\vec p}^{\: \prime}; 
    \lambda_1^\prime, \lambda_2^\prime\rangle
    \ \langle {\vec p}^{\: \prime}; 
    \lambda_1^\prime,\lambda_2^\prime  
    \vert \psi^\prime \rangle
\label{eq:r96a}\end {eqnarray}
is written in a rotationally covariant form.


\section{Rotational invariance}
\label{sec:3}

Integral equation~(\ref{eq:r96}) has rotationally covariant
but still not rotationally invariant form because of the interaction
kernel $\tilde{V}_{eff}$, written in the light-front frame.
Let us extract the part of interaction, which has manifestly
rotational symmetry.

Quite generally $\Theta$ factor  Eq.~(\ref{eq:r16}) is a function of the ratio
of its two arguments. Therefore, making use of Eq.~(\ref{eq:ii18b}),
the effective interaction Eq.~(\ref{eq:56}) is given as
\begin{eqnarray}
  V_{\rm eff} = - \frac{\alpha}{4\pi^2} 
                  \langle\gamma^\mu\gamma^\nu\rangle_{ex} 
	          B_{\mu\nu}
                - \frac{\alpha}{4\pi^2} 
                  \langle\gamma^\mu\gamma^\nu\rangle_{an} 
	          C_{\mu\nu}   
\,,\label{eq:inter}\end{eqnarray}
where the exchange part is defined
\begin{eqnarray}
 B_{\mu\nu} &=& \frac{g_{\mu\nu}}{Q^2}
  + \left( \frac{g_{\mu\nu}}{Q^2}
         - \frac{\eta_{\mu}\eta_{\nu}}{{q^+}^2} \right)
    \frac{\xi^2-\xi\vartheta(\xi)}{1-\xi^2}
\nonumber\\
 \vartheta(\xi) &=& \Theta_{e \bar{e}} - \Theta_{\bar{e} e}
 \quad,\quad
 \xi = \frac{\delta Q^2}{Q^2}
\,,\label{eq:eq1}\end{eqnarray}
and the annihilation part
\begin{eqnarray}
 C_{\mu\nu} &=& \frac{g_{\mu\nu}}{M^2}
  + \left( \frac{g_{\mu\nu}}{M^2} 
         - \frac{\eta_{\mu}\eta_{\nu}}{{p^+}^2} \right)
   \frac{\beta^2-\beta\chi(\beta)}{1-\beta^2}
\nonumber\\
 \chi(\beta) &=& \Theta_{a b} - \Theta_{b a}
 \quad,\quad
 \beta = \frac{\delta M^2}{M^2}
\,.\label{eq:eq2}\end{eqnarray}
The terms in the effective interaction proportional to
$\vartheta(\xi)$, $\chi(\beta)$ depend explicitly on the choice of
cut-off function and arise from $l$-ordering of the generator 
in the operator of unitary transform~\cite{previous}.
Explicit form of the effective interaction 
with different cut-off functions is given in Appendix A.


Define energy denominators in equation~(\ref{eq:eq2}).
Due to the three-momentum conservation on the light-front,
$p_1+p_2=p'_1+p'_2$ for longitudinal and transversal components,
one has $D_e-D_{\bar e}=D_a-D_b$, where the energy denominators $D_k$
in both channels are given in Eq.~(\ref{eq:40a}) and Eq.~(\ref{eq:40b}).
Therefore
\begin{eqnarray}
\delta Q^2 = \left( - \frac{q^+}{p^+} \right)\delta M^2
\,.\label{eq:delta}\end{eqnarray}
where $\delta M^2$ -- the total energy difference between initial
and final states shows the 'off-shellness` of process.

Using the parametrization Eq.~(\ref{eq:eq3})
one has for the energy denominators
\begin{eqnarray}
 Q^2 &=& {\vec q}^{\, 2} -p_z p'_z\frac{(M_a-M_b)^2}{M_aM_b}
\nonumber\\
 \delta Q^2 &=& - \left( \frac{p'_z}{M_a}
                       - \frac{p_z} {M_b} \right)\delta M^2
\nonumber\\
 M_a^2 &=& 4({\vec p}^{\: \prime 2}+m^2)
\nonumber\\
 M_b^2 &=& 4({\vec p}^{\: 2}+m^2)
\,,\label{eq:note}\end{eqnarray}
where $q=p'-p=(q_z,q_{\perp})$ is 
the three-momentum transfer of the photon,
and the relations between mean-squared and difference momenta
and corresponding energy differences are given in Eq.~(\ref{eq:r15})
and Eq.~(\ref{eq:r16}) for exchange and annihilation channels, respectively. 

The second term in Eq.~(\ref{eq:eq2}) is obviously not rotational invariant.
For the real processes, $\delta M^2=0$, the second term 
vanishes for both channels,
and the effective interaction is independent on the cutoff function
and coincide with the result of Tamm-Dancoff approximation  
\begin{eqnarray}
  V_{\rm eff} &=& -\frac{\alpha}{4\pi^2}
   \frac{\langle\gamma^\mu \gamma_\mu\rangle}{{\vec q}^{\,2}}
\,,\label{eq:td}\end{eqnarray}
and similarly in annihilation channel.
The same holds to the leading order of nonrelativistic expansion
${\vec p}^{\, 2}/m^2 \ll 1$ ~\cite{previous}.

Estimate current-current term in exchange channel, Eq.~(\ref{eq:r94}),
which defines nominator of the interaction Eq.~(\ref{eq:td}).
We work in the Lepage-Brodsky convention for the spinors~\cite{LeBr}
\begin{eqnarray}  
 u(p,\lambda) &=& \frac{2}{p^+}
    ( p^+ +\beta m + \vec{\alpha}_{\perp}\vec{p}_{\perp} )
      \Lambda_{+} \chi_{\lambda}
\nonumber\\
              &=& \frac{2}{\sqrt{N}}
    ( E +\beta m + \vec{\alpha}\vec{p} )
      \Lambda_{+} \chi_{\lambda}   
\,,\label{eq:spinor}\end{eqnarray}
and similarly for $v(p,\lambda)$ with the change $m\rightarrow -m$
and $\chi_{\lambda}\rightarrow \chi_{-\lambda}$ in the above formula. 
The second expression for spinor holds quite in general for the solution
of Dirac equation, where $p^0=E=\sqrt{{\vec p}^{\, 2}+m^2}$
and the normalization factor $N=E+p_z$. Here $\beta=\gamma^0$, 
$\vec{\alpha}=\gamma^0\vec{\gamma}$, $\Lambda_{+}=1/2(1+\alpha^3)$ 
is the projection operator ~\cite{bpp97}, and the spinor 
\begin{displaymath}
 \chi_{\lambda} = {\xi_{\lambda} \choose 0}
\,,\label{eq:spinor1}\end{displaymath}
is defined through the usual two-component spinors
\begin{displaymath}
 \xi_{\uparrow} ={1 \choose 0}
 \quad; \quad
 \xi_{\downarrow}{0 \choose 1}
\,,\label{eq:spinor2}\end{displaymath} 
Using the explicit representation for $\gamma$ matrices
and projection operators $\Lambda_{+}, \Lambda_{-}$
and relations between them ~\cite{bpp97},
one has
\begin{eqnarray}
        \bar{u}(p,\lambda)\gamma^0 u(p^\prime,\lambda^\prime) &=&
  \frac{1}{\sqrt{NN^\prime}}\xi_{\lambda}^+
     ( (E+p_z)(E^\prime+p_z^\prime) +(\vec p {\vec p}^{\: \prime})
        +i[\vec p  \times {\vec p}^{\: \prime}]\vec \sigma
\nonumber\\
   &+&   i[\vec p  \times \vec \sigma]_z(p_z^\prime+m)
        -i[{\vec p}^{\: \prime}\times \vec \sigma]_z(p_z+m)
        +m^2-p_zp_z^\prime
     ) \xi_{\lambda^\prime}
\nonumber\\
        \bar{u}(p,\lambda)\gamma^i u(p^\prime,\lambda^\prime) &=&
 \frac{1}{\sqrt{NN^\prime}}\xi_{\lambda}^+
     (  (E+p_z)  (p^{\prime\, i} 
      +i[{\vec p}^{\: \prime} \times \vec \sigma]^{i})
      + (E^\prime+p^\prime_z)(p^{i} -i[\vec p \times \vec \sigma]^{i})
\nonumber\\ 
   &+&  \delta^{iz} ( EE^\prime-m^2-(\vec p {\vec p}^{\: \prime})
      -i[\vec p  \times {\vec p}^{\: \prime}]\vec{\sigma}
\nonumber\\
   &+& i[\vec p  \times \vec \sigma]^{i}(E^\prime -m)
      -i[{\vec p}^{\: \prime}\times \vec \sigma]^{i}(E -m) ) 
\nonumber\\ 
   &+&   i\varepsilon^{ij}\sigma^{j}
       ( (E+p_z)(m+p_z^\prime)-(E^\prime+p_z^\prime)(m+p_z) )  
     ) \xi_{\lambda^\prime}
\,,\label{eq:a}\end{eqnarray}
where $i=1,2,3$; $p=(p_z,\vec{\kappa}_{\perp})$ is the three momentum and
$\varepsilon^{ij}=\varepsilon^{ijz}$, $\varepsilon^{12}=-\varepsilon^{21}=1$.
Introducing three-momentum transfer and its mean
\begin{eqnarray}
  \vec q &=& {\vec p}^{\: \prime}-\vec p
\nonumber\\
  \vec k &=& \frac{1}{2}({\vec p}^{\: \prime}+\vec p)
\,,\label{}\end{eqnarray}
where $4(\vec q\vec k)=\delta M^2=2(E^2-E^{\prime\, 2})$ 
with $\delta M^2$ defined above,
Eq.~(\ref{eq:a}) is written
\begin{eqnarray}
 \bar{u}(p,\lambda)\gamma^0 u(p^\prime,\lambda^\prime) &=&
  \frac{1}{\sqrt{NN^\prime}}\xi_{\lambda}^+
      ( (E+k_z)(E^\prime +k_z) +\vec k^2-{\vec q}^{\, 2}/4
         -i[\vec q \times \vec k]\vec \sigma
\nonumber\\
      &-& i[\vec q \times \vec \sigma]_z
          (k_z+m)
         +q_z((E-E^\prime)/2
         +i[\vec k \times \vec \sigma]_z)
         +m^2-k_z^2
      ) \xi_{\lambda^\prime}
\nonumber\\
 \bar{u}(p,\lambda)\gamma^i u(p^\prime,\lambda^\prime) &=&
 \frac{1}{\sqrt{NN^\prime}}\xi_{\lambda}^+
      ( ((E+E^\prime)/2+k_z)
        (2k^{i} +i[\vec q \times\vec \sigma]^{i})
\nonumber\\
    &+& ((E-E^\prime) -q_z)
        (q^{i}/2 +i[\vec k \times\vec \sigma]^{i})
\nonumber\\
    &+&  \delta^{iz} ( EE^\prime-m^2-(\vec{k}^2-{\vec q}^{\, 2}/4)
       +i[\vec q \times\vec k]\vec{\sigma}
\nonumber\\
    &-& i[\vec q\times \vec \sigma]^{i}
        ((E+E^\prime)/2-m)
       - i[\vec k \times\vec \sigma]^{i} 
        (E-E^\prime) ) 
\nonumber\\
    &+& i\varepsilon^{ij}\sigma^{j}
       ((k_z+m)(E-E^\prime)+q_z((E+E^\prime)/2-m))  
      ) \xi_{\lambda^\prime}
\,,\label{eq:b}\end{eqnarray}
where no approximations are done so far.
Excluding the overall normalization factor
the first lines in Eq.~(\ref{eq:b}) for scalar and vector current terms
contain rotationally invariant parts 
(except terms proportional to $k_z$), 
which coincide with the corresponding expressions
when making use of Bjorken-Drell convention for spinors ~\cite{MePa}.
\footnote{
Spinors as used by Bjorken-Drell are
\begin{eqnarray}  
 u(p,\lambda) &=& \frac{1}{\sqrt{N}}
    ( E +\beta m + \vec{\alpha}\vec{p} )
       \chi_{\lambda}   
\,,\label{eq:fn1}\end{eqnarray}
where $N=E+m$ and $\chi_{\lambda}$ is defined in the main text.
The correspoding expressions for current terms are
\begin{eqnarray}
 \bar{u}(p,\lambda)\gamma^0 u(p^\prime,\lambda^\prime) &=&
  \frac{1}{\sqrt{NN^\prime}}\xi_{\lambda}^+
     ( (E+m)(E^\prime+m) +\vec{k}^2-\vec{q}^{\, 2}/4
         -i[\vec q \times \vec k]\vec \sigma
     ) \xi_{\lambda^\prime}
\nonumber\\
 \bar{u}(p,\lambda)\gamma^i u(p^\prime,\lambda^\prime) &=&
 \frac{1}{\sqrt{NN^\prime}}\xi_{\lambda}^+
     ( ((E+E^\prime)/2+m)
        (2k^{i}  +i[\vec q \times \vec \sigma]^{i})
\nonumber\\
      &+& (E-E^\prime)
        (q^{i}/2 +i[\vec k \times \vec \sigma]^{i})
     ) \xi_{\lambda^\prime}
\,,\label{eq:fn2}\end{eqnarray}  
For the energy conserving process this expression was obtained
in ~\cite{MePa}.
}

Merkel et.al.~\cite{MePa} showed, that as far as the energy is conserved,
this part gives rise to familiar spin dependent forces.
The rest terms in Eq.~(\ref{eq:b}) are obviously not rotationally invariant,
particularly when the spacial rotations are performed
perpendicular to the $z$-axis. Expanding expression Eq.~(\ref{eq:b})
to the second order in $|\vec p|/m \ll 1$ and performing
the unitary transformation in spin space, Brisudova et.al.~\cite{BrPe2}
obtained Breit-Fermi spin-spin and tensor interactions.
It seems to be impossible to reproduce full set of Breit-Fermi terms
from the second order effective interaction in the light-front gauge.
Also it is complicated to cover rotational symmetry
on the level of light-front effective Hamiltonians 
without additional approximations are done.
We use directly the effective electron-positron 
interaction Eq.~(\ref{eq:inter})
for numerical calculations of positronium spectrum.
We aim to get fine structure and to investigate rotational symmetry
on the level of spectrum. The impact of different cutoff functions
is also considered. 
The results of these calculations are presented in the next section.

\begin{table}
\begin{tabular}{c||c||c||c||c||c||c}
$n$ & Term & $B_{ETPT}$ & $B_E$ & $B_G^{\eta}$& $B_G$ & $B_S$ \cr \hline \hline
1   & $1^1S_0$ & 1.118125   & 1.049550 & 1.101027 & 1.026170 & 0.920921 \\
2   & $1^3S_1$ & 0.998125   & 1.001010 & 1.049700 & 0.981969 & 0.885347 \\
3   & $2^1S_0$ & 0.268633   & 0.260237 & 0.266490 & 0.260642 & 0.242607 \\
4   & $2^3S_1$ & 0.253633   & 0.253804 & 0.259506 & 0.254765 & 0.234312 \\
5   & $2^1P_1$ & 0.253633   & 0.257969 & 0.263056 & 0.257664 & 0.237611 \\
6   & $2^3P_0$ & 0.261133   & 0.267070 & 0.273826 & 0.266563 & 0.243075 \\
7   & $2^3P_1$ & 0.255508   & 0.259667 & 0.265412 & 0.260127 & 0.238135 \\
8   & $2^3P_2$ & 0.251008   & 0.255258 & 0.260345 & 0.255498 & 0.236383 
\end{tabular}
\caption{Binding coefficients, $B_n=4 (2-M_n)/\alpha^2$
($\alpha=0.3$), for the lowest modes of the positronium spectrum 
at $J_z=0$ for the equal time perturbation theory up to order 
$\alpha^4$ ($B_{ETPT}$) compared to our
calculations with exponential ($B_E$), Gaussian ($B_G$) and sharp
($B_S$) cutoffs. $B_G$ is obtained using only $g_{\mu\nu}$ part
of interaction; for $B_G^{\eta}$ $'\eta_\mu\eta_\nu`$ term is included.   
Exchange channel is considered.}
\end{table}

\begin{table}
\begin{tabular}{c||c||c||c||c||c}
$n$ & Term     & $B_E$    & $B_G^{\eta}$& $B_G$    & $B_S$    \cr \hline \hline
1   & $1^1S_0$ & 1.049550 & 1.101270 & 1.026170 & 0.920921 \\
2   & $1^3S_1$ & 0.936800 & 0.978018 & 0.921847 & 0.834004 \\
3   & $2^1S_0$ & 0.260237 & 0.266490 & 0.260642 & 0.242624 \\
4   & $2^3S_1$ & 0.255292 & 0.260383 & 0.255615 & 0.234338 \\
5   & $2^1P_1$ & 0.257969 & 0.263056 & 0.257664 & 0.236383 \\
6   & $2^3P_0$ & 0.267090 & 0.273847 & 0.266626 & 0.243075 \\
7   & $2^3P_1$ & 0.259667 & 0.265412 & 0.260127 & 0.237611 \\
8   & $2^3P_2$ & 0.245615 & 0.250821 & 0.247091 & 0.230901  
\end{tabular}
\caption{Binding coefficients, $B_n=4 (2-M_n)/\alpha^2$
($\alpha=0.3$), for the lowest modes of the positronium spectrum 
at $J_z=0$ for our calculations with exponential ($B_E$), 
Gaussian ($B_G$) and sharp ($B_S$) cutoffs. 
$B_G^{\eta}$ includes $'\eta_\mu\eta_\nu`$ term
in exchange channel; $B_G$ does not. 
Exchange and annihilation channels are considered.}
\end{table}

\begin{table}
\begin{tabular}{c|| c|| c|| c|| c}
$n$ &Term & $\delta B_E$ & $\delta B_G$ & $\delta B_S$ \\ \hline \hline\\[-8pt]
2   & $1^3S_1$ &  6.30 $10^{-4}$  &  1.76 $10^{-3}$  & 1.18 $10^{-3}$ \\
4   & $2^3S_1$ &  8.40 $10^{-5}$  &  1.77 $10^{-4}$  & 9.0 $10^{-5}$ \\
5   & $2^1P_1$ & -1.30 $10^{-5}$  & -7.47 $10^{-4}$  &-9.1 $10^{-5}$ \\
7   & $2^3P_1$ & -4.08 $10^{-4}$  & -4.08 $10^{-4}$  & 1.4 $10^{-4}$ \\
8   & $2^3P_2$ &  5    $10^{-6}$  & -7.7 $10^{-5}$   & 4.15 $10^{-4}$ 
\end{tabular}
\caption{Difference in the corresponding energy levels between $J_z\!=\!0$
and $J_z\!=\!1$ states for exponential ($\delta B_E$), 
Gaussian ($\delta B_G$) and sharp ($\delta B_S$) cutoffs.
Exchange is channel is considered.}
\end{table}

\begin{table}
\begin{tabular}{c|| c|| c|| c|| c}
$n$ & Term & $\delta B_E$ & $\delta B_G$ & $\delta B_S$ \\ \hline \hline\\[-8pt]
2   & $1^3S_1$ & -1.411 $10^{-3}$ & -7.86 $10^{-4}$  &-1.65 $10^{-3}$ \\
4   & $2^3S_1$ & -4.1 $10^{-5}$   & -4.0 $10^{-5}$   &-1.15 $10^{-4}$ \\
5   & $2^1P_1$ & -6.4 $10^{-5}$   & -6.52 $10^{-4}$  &-4.60 $10^{-4}$ \\
7   & $2^3P_1$ & -4.69 $10^{-4}$  & -4.74 $10^{-4}$  &-1.40 $10^{-4}$ \\
8   & $2^3P_2$ & -1.96 $10^{-4}$  & -1.36 $10^{-4}$  &-2.44 $10^{-4}$ 
\end{tabular}
\caption{Difference in the corresponding energy levels between $J_z\!=\!0$
and $J_z\!=\!1$ states for exponential ($\delta B_E$), 
Gaussian ($\delta B_G$) and sharp ($\delta B_S$) cutoffs.
Exchange and annihilation channels are considered.}
\end{table}

\section{Mass spectrum of positronium}
\label{sec:6.3}

We solve the integral equation~(\ref{eq:r96}),
with interaction kernel given in Eq.(\ref{eq:56}), 
for positronium spectrum numerically.
Effective interaction with different choice of cutoffs
is sumarized in Appendix A.

In polar coordinates the light-front variables are
$(\vec \kappa_{\perp};x)=(\kappa_{\perp},\varphi;x)$;
therefore the matrix elements of the effective interaction Eq.(\ref{eq:56})
depend on the angles $\varphi$ and $\varphi^{\prime}$, i.e. 
$\langle x,\kappa_{\perp},\varphi;\lambda_1,\lambda_2|V_{\rm eff}|
x',\kappa'_{\perp},\varphi';\lambda'_1,\lambda'_2\rangle$.
In order to introduce the spectroscopic notation for positronium
mass spectrum we integrate out the angular degree of freedom,
$\varphi$, introducing a discrete quantum number
$J_z=n$, $n\in {\bf Z}$
(actually for the annihilation channel only $|J_z|\leq 1$ is possible),
\begin{eqnarray}
&&  \hspace{-1.5cm} 
      \langle x, \kappa_{\perp}; J_z, \lambda_1, \lambda_2
      |\tilde{V}_{\rm eff}|
       x',\kappa'_{\perp};J'_z,\lambda'_1,\lambda'_2\rangle
\nonumber\\
&=& \frac{1}{2\pi}\int_0^{2\pi}d\varphi {\rm e}^{-iL_z\varphi}
                  \int_0^{2\pi}d\varphi'{\rm e}^{iL'_z\varphi'}
      \langle x, \kappa_{\perp}, \varphi; \lambda_1, \lambda_2
      |V_{\rm eff}|
       x',\kappa'_{\perp},\varphi';\lambda'_1,\lambda'_2\rangle
\nonumber\\
&&
\,\label{eq:r43}\end{eqnarray}
where $L_z=J_z-S_z$; $S_z=\frac{\lambda_1}{2}+\frac{\lambda_2}{2}$ 
and the states
can be classified (strictly speaking only for rotationally invariant
systems) according to their quantum numbers of total angular momentum $J$,
orbit angular momentum $L$, and total spin $S$. 
Definition of angular momentum operators in light-front
dynamics is problematic because they include interactions.

The matrix elements of the effective interaction
before integrating over the angles,
$ \langle x, \kappa_{\perp}, \varphi; \lambda_1, \lambda_2
|V_{\rm eff}| x',\kappa'_{\perp},\varphi';\lambda'_1,\lambda'_2\rangle$,
and after the integration inroducing the total momentum, $J_z$,
$ \langle x, \kappa_{\perp}; J_z, \lambda_1, \lambda_2|\tilde{V}_{\rm eff}|
   x',\kappa'_{\perp};J'_z,\lambda'_1,\lambda'_2\rangle$
for different cutoff functions are given in the exchange and 
annihilation channels in Appendices B and C, respectively.

Now we proceed to solve for the positronium spectrum 
in all sectors of $J_z$. For this purpose we formulate 
the light-front integral equation Eq.~(\ref{eq:r96}) in the form
where the integral kernel is given by the effective interaction
for the total momentum $J_z$, Eq.~(\ref{eq:r43}). After the change 
of variables Eq.~(\ref{eq:eq3}) we parametrize
$ \vec{p}=(\vec{\kappa}_{\perp},p_z)=
(\mu\sin\theta\cos\varphi,\mu\sin\theta\sin\varphi,\mu\cos\theta)$.
The Jacobian of this transformation Eq.~(\ref{eq:eq4}) is given 
\begin{eqnarray}
&& J=\frac{1}{2}
\frac{m^2+\mu^2 \sin^2\theta}{(m^2+\mu^2)^{3/2}}
\,.\label{r47}\end{eqnarray}

\newpage
One obtaines then the integral equation
\begin{eqnarray}
  && (M_n^2-4(m^2+\mu^2))
   \tilde{\psi}_n(\mu,\cos\theta;J_z, \lambda_1, \lambda_2)
\nonumber\\
  &+&   \sum_{J'_z,\lambda'_1,\lambda'_2}
        \int_{D}d\mu'\int_{-1}^{+1}d\cos\theta'
        \frac{\mu^{'2}}{2}
        \frac{m^2+\mu^{'2}(1-\cos^2\theta')}{(m^2+\mu^{'2})^{3/2}}
\nonumber\\
 &\times& \langle\mu, \cos\theta; J_z, \lambda_1, \lambda_2
         |\tilde{V}_{\rm eff}|
               \mu',\cos\theta';J'_z,\lambda'_1,\lambda'_2\rangle
 \tilde{\psi}_n
        (\mu',\cos\theta';J'_z,\lambda'_1,\lambda'_2)=0
\,.\label{eq:r48}\end{eqnarray}
The integration domain $D$, defined in Eq.~(\ref{eq:domain}), 
is given now by $\mu\in [0;\frac{\Lambda}{2}]$.
Neither $L_z$ nor $S_z$ are good quantum numbers; therefore
we set $L_z=J_z-S_z$.

The integral equation Eq.~(\ref{eq:r48}) is used to calculate
positronium mass spectrum numerically.
Note, that if one succeeds to integrate out the angular degrees of freedom
for the effective interaction Eq.~(\ref{eq:r43}) analytically,
one has $2$-dimensional integration in Eq.~(\ref{eq:r48})
instead of $3$-dimensional one in the original
integral equation~(\ref{eq:r96}) to perform numerically.


We use the numerical code ~\cite{TrPa2}, worked out by Uwe Trittmann  
for the similar problem ~\cite{TrPa}. 
This code includes for the numerical integration
the Gauss-Legendre algorithm (Gaussian quadratures). 
To improve the numerical convergence
the technique of Coulomb counterterms is included.
The problem has been solved for all components of the total
angular momentum, $J_z$. 

Positronium spectrum is mainly defined by the Coulomb singularity
\begin{eqnarray}
 \vec q \longrightarrow 0
\,,\label{eq:limit1}\end{eqnarray}
which is an integrable one analytically and also, by use of technique 
of Coulomb counterterms, numerically. In this region
$\delta Q^2\rightarrow 0$ and the energy denominator
$Q^2\rightarrow {\vec q}^{\, 2}$ Eq.~(\ref{eq:note}),
giving rise to the leading order Coulomb behavior for the effective
interaction Eq.~(\ref{eq:td}), independent on the cutoff function.
We use therefore standard Coulomb counterterms,
introduced for the Coulomb problem Eq.~(\ref{eq:td}) ~\cite{TrPa,TrPa2},
in the case of all cutoffs. 
Basing also on the argument Eq.~(\ref{eq:limit1}), we expect
the same pattern of levels for different cutoffs,
that we prove numerically to be true.

Another important limiting case to study effective interaction 
Eq.~(\ref{eq:inter}), namely its exchange part, is the collinear limit
\begin{eqnarray}
  q^+ \longrightarrow 0
\,,\label{eq:limit2}\end{eqnarray}  
that is special for light-front calculations.
Because of Eq.~(\ref{eq:delta}) the variable $\xi^2 \sim q^{+\, 2}$,
resulting for the $'\eta_\mu\eta_\nu'$ part of effective interaction
to be 
\begin{eqnarray} 
 \frac{\eta_\mu\eta_\nu}{q^{+\, 2}}\xi^2(1-\vartheta^{\prime}(0))
\,,\label{eq:}\end{eqnarray}
which is finite in this limit. This is true for 
the regular cutoff functions, as in the case of exponential 
and gaussian cutoffs, where the derivative 
$d\vartheta(0)/d\xi$ is well defined. For sharp cutoff
this condition is not fulfilled, and the effective interaction
contains the $1/q^+$ type of singularity in this case
(see Appendix A). We do not associate any physics with this singularity,
considering it as a consequence of artificial choice of cutoff,
which corresponds to singular generator of unitary transformation
Eq.~(\ref{eq:gen}). We omit the $'\eta_\mu\eta_\nu'$ term
in exchange channel for sharp cutoff in numerical calculations.

We argued that the region of Coulomb singularity,
and hence $'g_{\mu\nu}'$ part of effective interaction,
determines mainly the positronium spectrum. However,
including $'\eta_\mu\eta_\nu'$ part for gaussian cutoff
shifts all levels as a whole of about $5-7\%$,
since this part is diagonal in spin space (Appendix B),
and improves the data to be near the result obtained in covariant
equal time calculations (Table $1$). Presumably,
it is necessary to take into account $'\eta_\mu\eta_\nu'$ term
in exchange channel also for sharp cutoff
after the proper regularization of infrared longitudinal
divergences is done.

We place the results of calculations for three different cutoffs,
performed in exchange and including both exchange and annihilation
channels, in Tables $1$ and $2$, respectively.   
The corresponding set of figures is presented in Fig.$1$
and Fig.$2$.
We get the ionization threshold at $M^2\sim 4m^2$,
the Bohr spectrum, and the fine structure.
Including annihilation part increases the splittings twice as large
for the lowest multiplets.

As one can see from presents figures, certain mass eigenvalues at 
$J_z=0$ are degenerate with certain eigenvalues at other $J_z$
to a very high degree of numerical precision. As an example,
consider the second lowest eigenvalue for $J_z=0$.
It is degenerate with the lowest eigenvalue for $J_z=\pm 1$,
and can thus be classified as a member of the triplet with $J=1$.
Correspondingly, the lowest eigenvalue for $J_z=0$ having no companion
can be classified as the singlet state with $J=0$.
Quite in general one can interpret degenerate multiplets 
as members of a state with total angular momentum
$J=2J_{z,max}+1$. One can get the quantum number
of total angular momentum $J$ from the number of degenerate states
for a fixed eigenvalue $M_n^2$. One can make
contact with the conventional classification scheme 
$^{2S+1}L_{J}^{J_z}$, as indicated in Tables $1-2$.

Such pattern of spectrum is driven by rotational invariance.
To trace rotational symmetry we calculate the difference
of energy levels between $J_z=0$ and $J_z=1$ states
for the lowest multiplets. The data are given for exchange and
including annihilation channnel in Tables $3$ and $4$, 
respectively.
Annihilation part makes corresponding states practically degenerate
(see Tables $4$ and Figure $2$).

\section{Conclusion}
\label{sec:6.4}

The numerical solution of positronium bound state problem,
with the effective electron-positron interaction obtained by the flow equations,
is presented. No approximations along numerical procedure are done.

Concerning the spin-splittings the best agreement with covariant calculations 
is obtained for gaussian cutoff, the worst results are for sharp cutoff.
Rotational invariance is traced on the level of spectrum by studing
the degree of degeneracy of corresponding states with the same total momentum
but different projection $J_z$ in the multiplet. Again, better results
are obtained for exponential and gaussian cutoff functions than for sharp cutoff.
This suggests, that smooth cutoff functions are preferable to perform calculations. 
Including annihilation channel improves the extend of degeneracy.

For the sharp cutoff the lowest multiplet is placed higher than the one 
in case of exponential and gaussian cutoffs. The reason is in disregarding
the infrared divergent part, which is diagonal in spin space and shifts
the spectrum as a whole down. The question how to regularize this part
and include it in mass spectrum calculations should be considered.
Generally, the impact of the different choice of cutoff functions
on the spectrum is small.

In this work we solve the bound state integral equation for the one fixed
integration interval. Integration domain introduces the ultraviolet-cutoff
dependence of invariant mass squared $M^2(\Lambda)$, that reflects
renormalization group properties of the effective coupling constant.
We leave this question for the future study.

\newpage
\begin{figure}[thbp]
\centerline{\epsfxsize=\textwidth \epsfbox{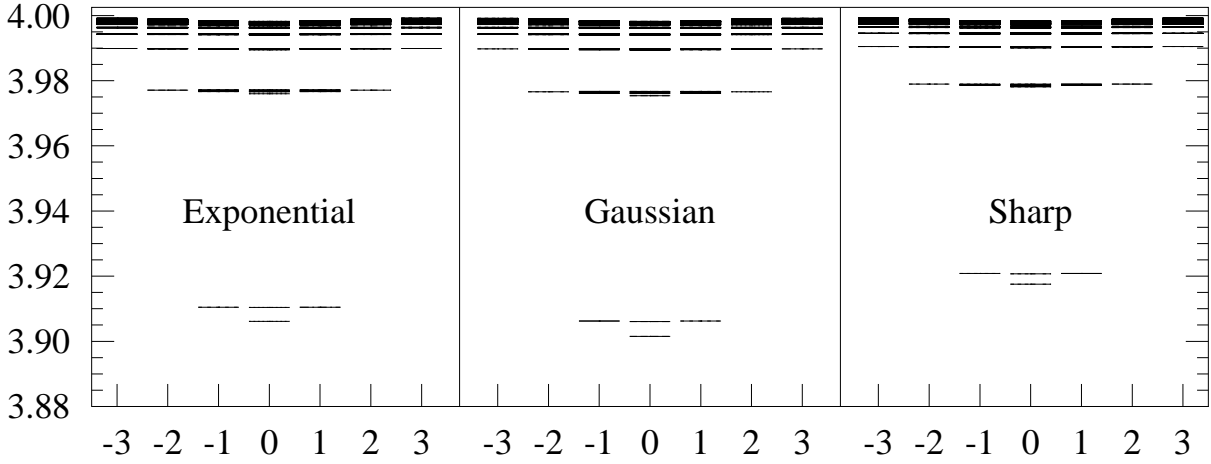}}
\caption{The invariant mass-squared spectrum $M_i^2$ for positronium 
   versus the projection of the total spin, $J_z$, excluding
   annihilation with exponential, Gaussian and sharp cutoffs.
   The number of integration points is $N_1=N_2=21$.}
\label{fig:1}
\end{figure}

\begin{figure}[thbp]
\centerline{\epsfxsize=\textwidth \epsfbox{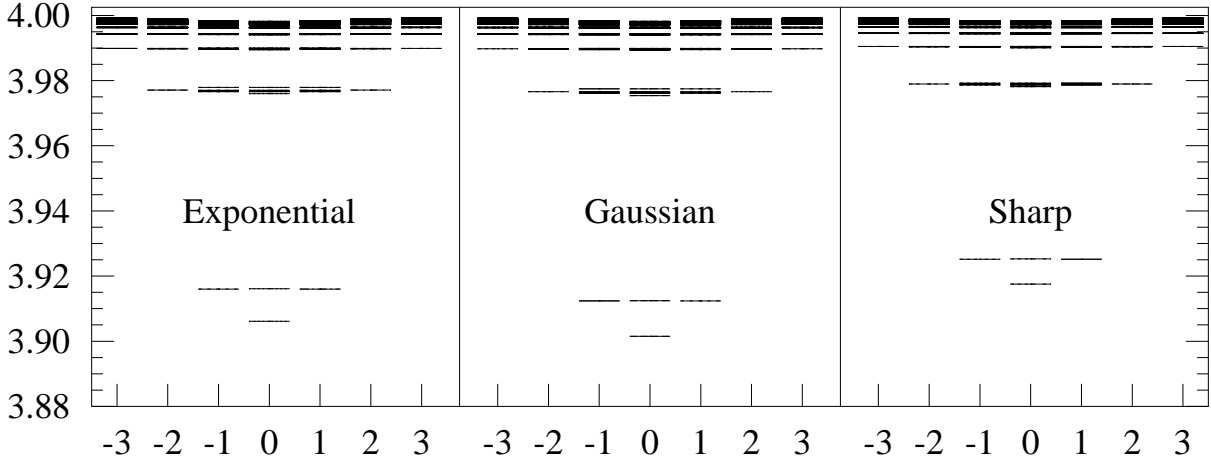}}
\caption{The invariant mass-squared spectrum $M_i^2$ for positronium 
   versus the projection of the total spin, $J_z$, including
   annihilation with exponential, Gaussian and sharp cutoffs.
   The number of integration points is $N_1=N_2=21$.}
\label{fig:2}
\end{figure}

\newpage
\appendix
\section{Defining different cut-offs}
\label{app:a}

In this appendix we summarize the results for 
the effective electron-positron interaction, generated 
by the flow equations with different similarity functions. 
In the practical work, three different similarity function
will be studied explicitly:
 
(1) the exponential cut-off, 
(2) the gaussian cut-off, and
(3) the sharp cut-off.

\noindent
(1) {\bf Exponential cut-off} 
\begin{eqnarray}
 f(D;l) &=& {\rm exp}\left(- \vert D \vert l \right)
\nonumber\\
 \Theta(D_e,D_{\bar{e}}) &=& \frac{D_e}{D_e+D_{\bar{e}}}
 ~;~ \vartheta(\xi) = \xi
\nonumber\\
 V_{\rm eff} &=& \frac{\alpha}{4\pi^2}
        \langle\gamma^{\mu}\gamma^{\nu}\rangle_{ex}
         \frac{g_{\mu\nu}}{q^+ D}
 -\frac{\alpha}{4\pi^2}
        \langle\gamma^{\mu}\gamma^{\nu}\rangle_{an}
    \frac{g_{\mu\nu}}{p^+ \tilde{D}}
\nonumber\\
 &=& -\frac{\alpha}{4\pi^2}
           \langle\gamma^{\mu}\gamma_{\mu}\rangle_{ex}
         \frac{1}{Q^2}
 -\frac{\alpha}{4\pi^2}
           \langle\gamma^{\mu}\gamma_{\mu}\rangle_{an}
    \frac{1}{M^2}
\,,\label{eq:a1}\end{eqnarray}
where $D=1/2(D_e+D_{\bar{e}})$ and $\tilde{D}=1/2(D_a+D_b)$.
The first choice of similarity function
gives exactly the result of perturbation theory.

\noindent
(2) {\bf Gaussian cut-off}
\begin{eqnarray}
 f(D;l) &=& {\rm exp}\left(- D^2 l \right)
\nonumber\\
 \Theta(D_e,D_{\bar{e}}) &=& \frac{D_e^2}{D_e^2+D_{\bar{e}}^2}
~;~ \vartheta(\xi) = \frac{2\xi}{1+\xi^2} 
\nonumber\\
 V_{\rm eff} &=& \frac{\alpha}{4\pi^2}
           \langle\gamma^{\mu}\gamma^{\nu}\rangle_{ex}
   \left[ \frac{g_{\mu\nu}}{q^+}\frac{D_e+D_{\bar{e}}}{D_e^2+D_{\bar{e}}^2}
 -\frac{\eta_{\mu}\eta_{\nu}}{2{q^+}^2} 
 \frac{(D_e-D_{\bar{e}})^2}{D_e^2+D_{\bar{e}}^2} \right]
\nonumber\\
 &-& \frac{\alpha}{4\pi^2}
           \langle\gamma^{\mu}\gamma^{\nu}\rangle_{an}
\left[ \frac{g_{\mu\nu}}{p^+}\frac{D_a+D_b}{D_a^2+D_b^2}
 -\frac{\eta_{\mu}\eta_{\nu}}{2{p^+}^2}
 \frac{(D_a-D_b)^2}{D_a^2+D_b^2} \right]
\nonumber\\
 = &-& \frac{\alpha}{4\pi^2}
          \langle\gamma^{\mu}\gamma^{\nu}\rangle_{ex}
 \left[ \frac{g_{\mu\nu}}{Q^2}+\frac{\eta_{\mu}\eta_{\nu}}{{q^+}^2}
\frac{\delta Q^4}{Q^4} \right]
\frac{Q^4}{Q^4+\delta Q^4}
\nonumber\\
 &-& \frac{\alpha}{4\pi^2}
           \langle\gamma^{\mu}\gamma^{\nu}\rangle_{an}
 \left[ \frac{g_{\mu\nu}}{M^2}-\frac{\eta_{\mu}\eta_{\nu}}{{p^+}^2}
\frac{\delta M^4}{M^4} \right]
\frac{M^4}{M^4+\delta M^4}
\,,\label{eq:a2}\end{eqnarray}
where we understand under $Q^4= (Q^2)^2$
and $\delta Q^4= (\delta Q^2)^2$ with $Q^2$ and $\delta Q^2$ defined in
Eq.~(\ref{eq:r15}).

\noindent
(3) {\bf Sharp cut-off}
\begin{eqnarray}
 f(D;l) &=& \theta \left(1 -\vert D \vert l \right)
\nonumber\\
 \Theta(D_e,D_{\bar{e}}) &=&
 \theta \left(\vert D_e \vert -\vert D_{\bar{e}} \vert \right)
~;~ \vartheta(\xi) = {\rm sign} (\xi)
\nonumber\\
 V_{\rm eff} &=& \frac{\alpha}{4\pi^2}
        \langle\gamma^{\mu}\gamma^{\nu}\rangle_{ex}
        \left[ \frac{g_{\mu\nu}}{q^+}
 \left( \frac{\theta(\vert D_e\vert-\vert D_{\bar{e}}\vert)}{D_e}
   + \frac{\theta(\vert D_{\bar{e}}\vert-\vert D_e\vert)}{D_{\bar{e}}} 
  \right)  \right.
\nonumber\\
   &&\left. - \frac{\eta_{\mu}\eta_{\nu}}{2{q^+}^2}(D_e-D_{\bar{e}})
 \left( \frac{\theta(\vert D_e\vert-\vert D_{\bar{e}}\vert)}{D_e}
      - \frac{\theta(\vert D_{\bar{e}}\vert-\vert D_e\vert)}{D_{\bar{e}}} 
 \right)    
           \right]
\nonumber\\
 &-& \frac{\alpha}{4\pi^2}
            \langle\gamma^{\mu}\gamma^{\nu}\rangle_{an}
        \left[ \frac{g_{\mu\nu}}{p^+}
 \left( \frac{\theta(\vert D_a\vert-\vert D_b\vert)}{D_a}
      + \frac{\theta(\vert D_b\vert-\vert D_a\vert)}{D_b} \right)\right.
\nonumber\\
   &&\left. - \frac{\eta_{\mu}\eta_{\nu}}{2{p^+}^2}(D_a-D_b)
 \left( \frac{\theta(\vert D_a\vert-\vert D_b\vert)}{D_a}
      - \frac{\theta(\vert D_b\vert-\vert D_a\vert)}{D_b} \right)    
           \right]
\nonumber\\
 = &-& \frac{\alpha}{4\pi^2}
        \langle\gamma^{\mu}\gamma^{\nu}\rangle_{ex}
 \left[ \frac{g_{\mu\nu}}{Q^2}+\frac{\eta_{\mu}\eta_{\nu}}{{q^+}^2}
\frac{\vert \delta Q^2 \vert}{Q^2} \right]
\frac{Q^2}{Q^2+\vert \delta Q^2 \vert}
\nonumber\\
 &-& \frac{\alpha}{4\pi^2}
         \langle\gamma^{\mu}\gamma^{\nu}\rangle_{an}
 \left[ \frac{g_{\mu\nu}}{M^2}-\frac{\eta_{\mu}\eta_{\nu}}{{p^+}^2}
\frac{\vert \delta M^2 \vert}{M^2} \right]
\frac{M^2}{M^2+\vert \delta M^2 \vert}
\,,\label{eq:a3}\end{eqnarray}
The motivation to choose these cutoff functions is the following.
Using exponential cutoff in flow equations one generates
the same interaction as obtained also in Tamm-Dancoff approach,where
numerical calculations of positronium spectrum
are performed \cite{TrPa}, and we use this numerical code here.
Note also, that for this cutoff the effective interaction
looks very much as in covariant calculations: 
it contains only $'g_{\mu\nu}`$ part, and $'\eta_\mu\eta_\nu`$
part is identically zero, so that there is no collinear problem.
Gaussian cutoff corresponds to the original choice of generator
Eq.~(\ref{eq:gen}) by Wegner as commutator of diagonal, 
particle number conserving, and off-diagonal, particle number changing, 
parts of Hamiltonian.
Sharp cutoff is used often in the alternative
similarity scheme to perform calculations \cite{GlWi}.

\newpage
\section{The matrix elements in the exchange channel}
\label{app:b}

In this Appendix we follow the scheme of the work \cite{TrPa}
to calculate the matrix elements of the effective interaction
in the exchange channel.\footnote{
Some of these calculations can be found in ~\cite{gub98}. }
Here, we list the general, angle-dependent matrix elements
defining the effective interaction in the exchange channel 
and the corresponding matrix elements
of the effective interaction for arbitrary $J_z$, 
after integrating out the angles.
Exchange part of the effective interaction for three different
cut-offs Eqs.~(\ref{eq:a1}--\ref{eq:a3}) can be written
\begin{eqnarray}
  V_{\rm eff} = - \frac{\alpha}{4\pi^2} 
  \langle\gamma^\mu\gamma^\nu\rangle B_{\mu\nu}   
\,,\end{eqnarray}
where explicitly one has 

\noindent
(1) {\bf Exponential cut-off} 
\begin{eqnarray}
 B_{\mu\nu} = \frac{g_{\mu\nu}}{Q^2}
\,,\label{eq:b1}\end{eqnarray}

\noindent
(2) {\bf Gaussian cut-off}
\begin{eqnarray}
 B_{\mu\nu} = g_{\mu\nu}{\rm Re}\left( \frac{1}{Q^2+i\delta Q^2} \right)
            -\frac{\eta_{\mu}\eta_{\nu}}{{q^+}^2}\delta Q^2
                        {\rm Im}\left( \frac{1}{Q^2+i\delta Q^2} \right)
\,,\label{eq:b2}\end{eqnarray}
where ${\rm Re}$ and ${\rm Im}$ are real and imaginary parts, respectively.
and $i^2=-1$.

\noindent
(3) {\bf Sharp cut-off}
\begin{eqnarray}
  B_{\mu\nu} &=& g_{\mu\nu}
  \left( \frac{\theta(-\delta Q^2)}{Q^2-\delta Q^2}
      +  \frac{\theta( \delta Q^2)}{Q^2+\delta Q^2} \right)
\nonumber\\
             &-& \frac{\eta_{\mu}\eta_{\nu}}{{q^+}^2} \delta Q^2
  \left( \frac{\theta(-\delta Q^2)}{Q^2-\delta Q^2}
      -  \frac{\theta( \delta Q^2)}{Q^2+\delta Q^2} \right)
\,,\label{eq:b3}\end{eqnarray}
where $q=p'_1-p_1$ is the momentum transfer; and
$\langle\gamma^\mu\gamma^\nu\rangle$ for the exchange channel 
is given in Eq.~(\ref{eq:r94}). 
We omit index $`ex'$ everywhere. 

It is convenient to extract the angular dependence in the functions
\begin{eqnarray}
 Q_e^2         &=& a_1 - b\cos t
\nonumber\\
 Q_{\bar{e}}^2 &=& a_2 - b\cos t
\nonumber\\
        t      &=& \varphi-\varphi^{'}
\,,\end{eqnarray}
where we define
 \begin{eqnarray}
 \vec{k}_{\perp} &=& k_{\perp}(\cos\varphi,\sin\varphi)
\end{eqnarray}
in polar system; here the terms are given 
\begin{eqnarray}
 a_1 &=& \frac{x'}{x}k_{\perp}^2+\frac{x}{x'}k_{\perp}^{'2}
+m^2\frac{(x-x')^2}{xx'}
 \nonumber\\
&=& k_{\perp}^2+k_{\perp}^{'2}
 +(x-x')\left(k_{\perp}^2(-\frac{1}{x})-k_{\perp}^{'2}(-\frac{1}{x'})\right)
 +m^2\frac{(x-x')^2}{xx'}
\nonumber\\
 a_2 &=& \frac{1-x'}{1-x}k_{\perp}^2+\frac{1-x}{1-x'}k_{\perp}^{'2}
 +m^2\frac{(x-x')^2}{(1-x)(1-x')}
\nonumber\\
&=& k_{\perp}^2+k_{\perp}^{'2}
 +(x-x')\left(k_{\perp}^2\frac{1}{1-x}-k_{\perp}^{'2}\frac{1}{1-x'}\right)
 +m^2\frac{(x-x')^2}{(1-x)(1-x')}
\nonumber\\
 b &=& 2k_{\perp}k_{\perp}^{'}
\label{eq:b4}\end{eqnarray}
Then the functions in Eqs.~(\ref{eq:b1}--\ref{eq:b3}) are given
\begin{eqnarray}
 Q^2 &=& a - b\cos t
\nonumber\\
 \delta Q^2 &=& \delta a
\,,\end{eqnarray}
where
\begin{eqnarray}
       a &=& \frac{1}{2}(a_1+a_2)
\nonumber\\
\delta a &=& \frac{1}{2}(a_1-a_2)
\,,\end{eqnarray}
It is useful to display the matrix elements of the effective interaction
in the form of tables. The matrix elements depend on the one hand
on the momenta of the electron and positron, respectively, and on the other
hand on their helicities before and after the interaction.
The dependence on the helicities occur during the calculation
of these functions
$E(x,\vec{k}_{\perp};\lambda_1,\lambda_2|x',\vec{k}'_{\perp};
\lambda'_1,\lambda'_2)$
in part I and 
$G(x,k_{\perp};\lambda_1,\lambda_2|x',k'_{\perp};
\lambda'_1,\lambda'_2)$ in part II
as different Kronecker deltas \cite{LeBr}.
These functions are displayed in the form of helicity tables.
We use the following notation for the elements of the tables
\begin{eqnarray}
&& F_i(1,2)~\rightarrow ~E_i(x,\vec{k}_{\perp};x',\vec{k}'_{\perp});~
G_i(x,k_{\perp};x',k'_{\perp})
\label{eq:a4}\end{eqnarray}
Also we have used in both cases for the permutation of particle and
anti-particle
\begin{eqnarray}
&&F_3^{*}(x,\vec{k}_{\perp};x',\vec{k}'_{\perp})
=F_3(1-x,-\vec{k}_{\perp};1-x',-\vec{k}'_{\perp})
\label{eq:a5}\end{eqnarray}
one has the corresponding for the elements of arbitrary $J_z$;
in the case when the function additionally depends
on the component of the total angular momentum $J_z=n$
we have introduced
\begin{eqnarray}
&& \tilde{F}_i(n)=F_i(-n)
\label{eq:a6}\end{eqnarray}

\subsection{The helicity table}

To calculate the matrix elements of the effective interaction
in the exchange channel we use the matrix elements of the Dirac spinors
listed in Table $1$ \cite{LeBr}. 
Also the following holds
$\bar{v}_{\lambda'}(p)\gamma^{\alpha}v_{\lambda}(q)
=\bar{u}_{\lambda}(q)\gamma^{\alpha}u_{\lambda'}(p)$.

\begin{table}[htbp]
\begin{tabular}{|c||c|}
\hline
\parbox{1.5cm}{ \[\cal M\] } & 
\parbox{12cm}{
\[
\frac{1}{\sqrt{k^+k^{'+}}}
\bar{u}(k',\lambda') {\cal M} u(k,\lambda)
\]
}
\\\hline\hline
$\gamma^+$ & \parbox{12cm}{
\[
\hspace{2cm}
2\delta^{\lambda}_{\lambda'}
\]}
\\\hline
$\gamma^-$ &
\parbox{12cm}{
\[
\frac{2}{k^+k^{'+}}\left[\left(m^2+
k_{\perp}k'_{\perp}e^{+i\lambda(\varphi-\varphi')}\right)
\delta^{\lambda}_{\lambda'}
-m\lambda\left(k'_{\perp}e^{+i\lambda \varphi'}-
k_{\perp} e^{+i\lambda\varphi}\right)
\delta^{\lambda}_{-\lambda'}\right]
\]}
\\\hline
$\gamma_{\perp}^1$ & 
\parbox{12cm}{
\[
\left(\frac{k'_{\perp}}{k^{'+}}e^{-i\lambda\varphi'}+
\frac{k_{\perp}}{k^+}e^{+i\lambda\varphi}
\right)\delta^{\lambda}_{\lambda'}
+m\lambda\left(\frac{1}{k^{'+}}-\frac{1}{k^+}
\right)\delta^{\lambda}_{-\lambda'}
\]}
\\\hline
$\gamma_{\perp}^2$ &
\parbox{12cm}{
\[ 
i\lambda\left(\frac{k'_{\perp}}{k^{'+}}e^{-i\lambda\varphi'}-
\frac{k_{\perp}}{k^+}e^{+i\lambda\varphi}
\right)\delta^{\lambda}_{\lambda'}
+im\left(\frac{1}{k^{'+}}-\frac{1}{k^+}
\right)\delta^{\lambda}_{-\lambda'}
\]}\\
\hline
\end{tabular}
\caption[Matrix elements of the Dirac spinors]
{Matrix elements of the Dirac spinors.}
\end{table}

We introduce for the matrix elements entering in the effective
interaction Eqs.~(\ref{eq:b1}--\ref{eq:b3})
\begin{eqnarray}
   2E^{(1)}(x,\vec{k}_{\perp};\lambda_1,\lambda_2|
   x',\vec{k}_{\perp}^{'};\lambda_1^{'},\lambda_2^{'})
   &=& \langle\gamma^{\mu}\gamma^{\nu}\rangle g_{\mu\nu}
\,,\end{eqnarray}
with $ \langle\gamma^{\mu}\gamma^{\nu}\rangle g_{\mu\nu}
   = \frac{1}{2}\langle\gamma^+\gamma^-\rangle
   + \frac{1}{2}\langle\gamma^-\gamma^+\rangle
   - \langle\gamma_1^2\rangle-\langle\gamma_2^2\rangle$ 
and
\begin{eqnarray}
   2E^{(2)}(x,\vec{k}_{\perp};\lambda_1,\lambda_2|
   x',\vec{k}_{\perp}^{'};\lambda_1^{'},\lambda_2^{'})
   &=& \langle\gamma^{\mu}\gamma^{\nu}\rangle 
     \eta_{\mu}\eta_{\nu}\frac{1}{{q^+}^2} 
\,,\end{eqnarray}
with $\langle\gamma^{\mu}\gamma^{\nu}\rangle 
    \eta_{\mu}\eta_{\nu}
   = \langle\gamma^{+}\gamma^{+}\rangle $; 
where
\begin{eqnarray}
   \langle\gamma^{\mu}\gamma^{\nu}\rangle = 
   {(\bar{u}(x,\vec{k}_{\perp};\lambda_1)\gamma^{\mu}
   u(x',\vec{k}_{\perp}^{'};\lambda_1^{'}))
   \over \sqrt{xx'}}
   {(\bar{v}(1-x',-\vec{k}_{\perp}^{'};\lambda_2^{'})
   \,\gamma^{\nu}
   v(1-x,-\vec{k}_{\perp};\lambda_2))
   \over \sqrt{(1-x)(1-x')}}
\label{eq:a8}\end{eqnarray}

These functions are displayed in
Table~\ref{GeneralHelicityTable}.

\begin{table}[htb]
\begin{tabular}{|c||c|c|c|c|}\hline
\rule[-3mm]{0mm}{8mm}{\bf final : initial} & 
$(\lambda_1',\lambda_2')=\uparrow\uparrow$ 
& $(\lambda_1',\lambda_2')=\uparrow\downarrow$ 
& $(\lambda_1',\lambda_2')=\downarrow\uparrow$ &
$(\lambda_1',\lambda_2')=\downarrow\downarrow$ \\ \hline\hline
\rule[-3mm]{0mm}{8mm}$(\lambda_1,\lambda_2)=\uparrow\uparrow$ & $E_1(1,2)$  
& $E_3^*(1,2)$ & $E_3(1,2)$ & $0$ \\ \hline
\rule[-3mm]{0mm}{8mm}$(\lambda_1,\lambda_2)=\uparrow\downarrow$ & 
$E_3^*(2,1)$ & $E_2(1,2)$ & $E_4(1,2)$ 
& $-E_3(2,1)$ \\ \hline
\rule[-3mm]{0mm}{8mm}$(\lambda_1,\lambda_2)=\downarrow\uparrow$& $E_3(2,1)$ 
& $E_4(1,2)$ & $E_2(1,2)$  & 
$-E_3^*(2,1)$\\ \hline
\rule[-3mm]{0mm}{8mm}$(\lambda_1,\lambda_2)=\downarrow\downarrow$ & $0$ 
& $-E_3(1,2)$ & $-E_3^*(1,2)$ & 
$E_1(1,2)$\\ \hline
\end{tabular}
\caption{General helicity table defining the effective interaction
in the exchange channel.}
\label{GeneralHelicityTable}
\end{table}

The matrix elements 
$E_i^{(n)}(1,2)=E_i^{(n)}(x,\vec{k}_{\perp};x',\vec{k}'_{\perp})$
with $n=1$ and $n=2$ for $`g_{\mu\nu}'$ and $\eta_{\mu}\eta_{\nu}$ terms, 
respectively, are the following
\begin{eqnarray}
  E_1^{(1)}(x,\vec{k}_{\perp};x',\vec{k}_{\perp}^{'})
&=& m^2\left(\frac{1}{xx'}+\frac{1}{(1-x)(1-x')}\right)
    +\frac{k_{\perp}k_{\perp}^{'}}{xx'(1-x)(1-x')}
    {\rm e}^{-i(\varphi-\varphi^{'})}\nonumber\\
  E_2^{(1)}(x,\vec{k}_{\perp};x',\vec{k}_{\perp}^{'})
&=& m^2\left(\frac{1}{xx'}+\frac{1}{(1-x)(1-x')}\right)
    +k_{\perp}^2\frac{1}{x(1-x)}+k_{\perp}^{'2}\frac{1}{x'(1-x')}\nonumber\\
&+& k_{\perp}k_{\perp}^{'}
    \left(\frac{{\rm e}^{i(\varphi-\varphi^{'})}}{xx'}
    +\frac{{\rm e}^{-i(\varphi-\varphi^{'})}}{(1-x)(1-x')}\right)\nonumber\\
  E_3^{(1)}(x,\vec{k}_{\perp};x',\vec{k}_{\perp}^{'})
&=& -\frac{m}{xx'}
    \left(k_{\perp}^{'}{\rm e}^{i\varphi^{'}}
    -k_{\perp}\frac{1-x'}{1-x}{\rm e}^{i\varphi}\right)\nonumber\\
  E_4^{(1)}(x,\vec{k}_{\perp};x',\vec{k}_{\perp}^{'})
&=& -m^2\frac{(x-x')^2}{xx'(1-x)(1-x')}
\,,\label{eq:a9}\end{eqnarray}
and
\begin{eqnarray}
  E_1^{(2)}(x,\vec{k}_{\perp};x',\vec{k}_{\perp}^{'})
&=& E_2^{(2)}(x,\vec{k}_{\perp};x',\vec{k}_{\perp}^{'})
 =  \frac{2}{(x-x')^2}
\nonumber\\
  E_3^{(2)}(x,\vec{k}_{\perp};x',\vec{k}_{\perp}^{'})
&=& E_4^{(2)}(x,\vec{k}_{\perp};x',\vec{k}_{\perp}^{'})
 =  0
\,.\label{eq:ab9}\end{eqnarray}

\subsection{The helicity table for arbitrary $J_z$.}

Following the description given in the main text Eq.~(\ref{eq:r43})
we integrate out the angles in the effective interaction 
in the exchange channel.
For the matrix elements of the effective interaction
for an arbitrary $J_z=n$ with $n\in {\bf Z}$ we introduce the functions
$G(x,k_{\perp};\lambda_1,\lambda_2|x',k'_{\perp};\lambda'_1,\lambda'_2)=
\langle x,k_{\perp};J_z,\lambda_1,\lambda_2|\tilde{V}_{\rm eff}|
x',k'_{\perp};J'_z,\lambda'_1,\lambda'_2\rangle$ 
in the exchange channel
and obtain the helicity Table~\ref{HelicityTableJz}.

\begin{table}[htb]
\begin{tabular}{|c||c|c|c|c|}\hline
\rule[-3mm]{0mm}{8mm}{\bf final : initial} 
&$(\lambda'_1,\lambda'_2)=\uparrow\uparrow$
&$(\lambda'_1,\lambda'_2)=\uparrow\downarrow$
&$(\lambda'_1,\lambda'_2)=\downarrow\uparrow$
&$(\lambda'_1,\lambda'_2)=\downarrow\downarrow$\\\hline\hline
\rule[-3mm]{0mm}{8mm}$(\lambda_1,\lambda_2)=\uparrow\uparrow$
&$G_1(1,2)$&$G_3^*(1,2)$&$G_3(1,2)$&$0$\\\hline
\rule[-3mm]{0mm}{8mm}$(\lambda_1,\lambda_2)=\uparrow\downarrow$
&$G_3^*(2,1)$&$G_2(1,2)$&$G_4(1,2)$&$-\tilde{G}_3(2,1)$\\\hline
\rule[-3mm]{0mm}{8mm}$(\lambda_1,\lambda_2)=\downarrow\uparrow$
&$G_3(2,1)$&$G_4(1,2)$&$\tilde{G}_2(1,2)$&$-\tilde{G}_3^*(2,1)$\\\hline
\rule[-3mm]{0mm}{8mm}$(\lambda_1,\lambda_2)=\downarrow\downarrow$&$0$&$-
\tilde{G}_3(1,2)$&$-\tilde{G}_3^*(1,2)$&
$\tilde{G}_1(1,2)$\\
\hline
\end{tabular}
\caption[Helicity table of the effective interaction
in the exchange channel for arbitrary $J_z = \pm n$, $x>x'$.]
{\protect\label{HelicityTableJz}
Helicity table of the effective interaction
for $J_z = \pm n$, $x>x'$.}
\end{table}

Here, the functions $G_i(1,2)=G_i(x,k_{\perp};x',k'_{\perp})$
are given
\begin{eqnarray}
 G_1(x,k_{\perp};x',k_{\perp}^{'}) &=& 
    \left( \frac{m^2}{xx'}+\frac{m^2}{(1-x)(1-x')} \right)Int(|1-n|)
\nonumber\\
&+& \frac{k_{\perp}k_{\perp}^{'}}{xx'(1-x)(1-x')}Int(|n|)
 -  \frac{2\delta a}{(x-x')^2}\tilde{Int}(|1-n|)
\nonumber\\
 G_2(x,k_{\perp};x',k_{\perp}^{'}) &=& \left(
    m^2\left(\frac{1}{xx'}+\frac{1}{(1-x)(1-x')} \right)
 +  \frac{k_{\perp}^2}{x(1-x)}+\frac{k_{\perp}^{'2}}{x'(1-x')} 
    \right)Int(|n|)
\nonumber\\
&+& k_{\perp}k_{\perp}^{'}\left( \frac{1}{xx'}Int(|1-n|)
 +  \frac{1}{(1-x)(1-x')}Int(|1+n|) \right)
\nonumber\\
&-& \frac{2\delta a}{(x-x')^2}\tilde{Int}(|n|)
\nonumber\\
 G_3(x,k_{\perp};x',k_{\perp}^{'}) &=&
 -  \frac{m}{xx'}\left(
    k_{\perp}^{'}Int(|1-n|)
 -  k_{\perp}\frac{1-x'}{1-x}Int(|n|) 
    \right)
\nonumber\\
 G_4(x,k_{\perp};x',k_{\perp}^{'}) &=&
 -  m^2\frac{(x-x')^2}{xx'(1-x)(1-x')}Int(|n|)
\label{eq:a11}\end{eqnarray}
we define
\begin{eqnarray}
 I(n;a,b) = -\frac{\alpha}{2\pi^2} 
 \int_0^{2\pi}dt\frac{\cos nt}{a-b\cos t}
\,,\end{eqnarray}
then in Eq.~(\ref{eq:a11}) the following functions are introduced

\noindent
(1) {\bf Exponential cut-off} 
\begin{eqnarray}
 Int(n) &=& I(n;a,b)
\nonumber\\
 \tilde{Int}(n) &=& 0
\,,\label{eq:d1}\end{eqnarray}

\noindent
(2) {\bf Gaussian cut-off}
\begin{eqnarray}
 Int(n)         &=& {\rm Re}I(n;a+i\delta a,b)
\nonumber\\
 \tilde{Int}(n) &=& {\rm Im}I(n;a+i\delta a,b)
\,,\label{eq:d2}\end{eqnarray}

\noindent
(3) {\bf Sharp cut-off}
\begin{eqnarray}
         Int(n) &=& \theta(-\delta a)I(n;a-\delta a,b)
                 +  \theta( \delta a)I(n;a+\delta a,b)
\nonumber\\
 \tilde{Int}(n) &=& \theta(-\delta a)I(n;a-\delta a,b)
                 -  \theta( \delta a)I(n;a+\delta a,b)
\,,\label{eq:d3}\end{eqnarray}
also $a+\delta a =a_1$ and$a-\delta a =a_2$.  

Explicitly is used 
\begin{eqnarray}
 && \int_0^{2\pi}dt\frac{\cos nt}{a-b\cos t}
 =  2\pi\frac{1}{\sqrt{a^2-b^2}}
    \left( \frac{a-\sqrt{a^2-b^2}}{b} \right)^n
\nonumber\\
 && \int_0^{2\pi}dt\frac{\sin nt}{a-b\cos t}
 =  0
\,,\label{eq:int}\end{eqnarray}
where $a$ can contain imaginary part 
as in the case of gaussian cutoff.
%

\newpage
\section{The matrix elements in the annihilation channel}
\label{app:c}

We repeat the same calculations for the matrix elements 
of the effective interaction in the annihilation channel. 
Annihilation part of the effective interaction 
can be written
\begin{eqnarray}
  V_{\rm eff} = - \frac{\alpha}{4\pi^2} 
  \langle\gamma^\mu\gamma^\nu\rangle C_{\mu\nu}   
\,,\end{eqnarray}
where one has
\begin{eqnarray}
   C_{\mu\nu} &=& 
   g_{\mu\nu}^{\perp}
   \left(\frac{\Theta_{ab}} {M_a^2} +
         \frac{\Theta_{ba}} {M_b^2} \right) - 
   \frac{\eta_{\mu}\eta_{\nu}}{{p^+}^2}
\nonumber\\
              &=& \frac{g_{\mu\nu}^{\perp}}{M^2}\
   \frac{1-\beta\chi(\beta)}{1-\beta^2}
  -\frac{\eta_{\mu}\eta_{\nu}}{{p^+}^2}
\,,\end{eqnarray}
in the frame $p_{\perp}=0$.\footnote{
Indeed
$\langle\gamma^\mu\gamma^\nu\rangle g_{\mu\nu}
=\frac{1}{2}\langle\gamma^+\gamma^-\rangle
+\frac{1}{2}\langle\gamma^-\gamma^+\rangle 
+\langle\gamma^\mu\gamma^\nu\rangle 
g_{\mu\nu}^{\perp}$;
therefore it holds
\begin{eqnarray}
  g_{\mu\nu} = g_{\mu\nu}^{\perp}
+ \frac{\eta_\mu(p_\nu-p_\nu^{\perp})+\eta_\nu(p_\mu-p_\mu^{\perp})}{p^+}
- \frac{{p^{\perp}}^2}{{p^+}^2}\eta_\mu\eta_\nu
\nonumber 
\,,\end{eqnarray}
The $4$-momentum of the photon $p_{\mu}$ in the $t$-channel can be written
$p_{\mu}=p^{'}_{1\mu}+p^{'}_{2\mu}-\eta_{\mu}D_a/2
        =p_{1\mu}    +p_{2\mu}    -\eta_{\mu}D_b/2$ with $D_a,D_b$ defined
in Eq.~(\ref{eq:40b}).
The Dirac equation $(p_1+p_2)_{\mu}\bar{u}(p_1)\gamma^{\mu}v(p_2)=0$ 
allows then to write 
$p_{\mu}\bar{u}(p_1,\lambda_1)\gamma^{\mu}v(p_2,\lambda_2)
 =   -M_b^2/(2p^+)\eta_{\mu}              
        \bar{u}(p_1,\lambda_1)\gamma^{\mu}v(p_2,\lambda_2)$.
Thus, when $p_{\perp}=0$, one has
\begin{eqnarray} 
g_{\mu\nu} \rightarrow g_{\mu\nu}^{\perp}-\frac{\eta_\mu\eta_\nu}{{p^+}^2}M^2
\nonumber
\,,\end{eqnarray}
where the arrow means that this tensor should be contracted with 
$\langle\gamma^\mu\gamma^\nu\rangle$ in the annihilation channel.
}
Explicitly the annihilation part of the effective interaction
for different cut-offs Eq.~({\ref{eq:a1})- Eq.~(\ref{eq:a3}) is given

\noindent
(1) {\bf Exponential cut-off} 
\begin{eqnarray}
 C_{\mu\nu} = \frac{g_{\mu\nu}^{\perp}}{M^2}
\,,\label{eq:f1}\end{eqnarray}

\noindent
(2) {\bf Gaussian cut-off}
\begin{eqnarray}
 C_{\mu\nu} = g_{\mu\nu}^{\perp}\frac{M^2}{M^4+\delta M^4}
            - \frac{\eta_{\mu}\eta_{\nu}}{{p^+}^2}
\,,\label{eq:f2}\end{eqnarray}
             
\noindent
(3) {\bf Sharp cut-off}
\begin{eqnarray}
  C_{\mu\nu} = g_{\mu\nu}^{\perp}\left( 
               \frac{\theta(M_a^2-M_b^2)}{M_a^2}
             + \frac{\theta(M_b^2-M_a^2)}{M_b^2} \right)       
             - \frac{\eta_{\mu}\eta_{\nu}}{{p^+}^2}
\,\label{eq:f3}\end{eqnarray}
where
$p^+=p_1^++p_2^+$ is the total momentum; and 
$\langle\gamma^\mu\gamma^\nu\rangle$
for annihilation is defined in Eq.~(\ref{eq:r94}).
The functions present in Eq.~(\ref{eq:f1})- Eq.~(\ref{eq:f3}) 
are given in the light-front frame 
\begin{eqnarray}
 M_a^2 &=& \frac{{k'_{\perp}}^2+m^2}{x'(1-x')}
\nonumber\\
 M_b^2 &=&  \frac{k_{\perp}^2+m^2}{x(1-x)}
\,,\label{eq:a12}\end{eqnarray}
we remind also
\begin{eqnarray}
 M^2 &=& \frac{1}{2}(M_a^2+M_b^2)
\nonumber\\
 \delta M^2 &=& \frac{1}{2}(M_a^2-M_b^2)
\label{eq:a13}\end{eqnarray}

Note that the energy denominators of the effective interaction
in the annihilation channel 
do not depend on the angles $\varphi,\varphi'$.

\begin{table}[htb]
\begin{tabular}{|c||c|}
\hline
\parbox{1.5cm}{ \[\cal M\] } & 
\parbox{12.5cm}{
\[
\frac{1}{\sqrt{k^+k^{'+}}}
\bar{v}(k',\lambda') {\cal M} u(k,\lambda)
\]
}
\\\hline\hline
$\gamma^+$ & \parbox{12.5cm}{
\[
\hspace{2cm}
2\delta^{\lambda}_{-\lambda'}
\]}
\\\hline
$\gamma^-$ &
\parbox{12.5cm}{
\[
\frac{2}{k^+k^{'+}}\left[-\left(m^2-
k_{\perp}k'_{\perp}e^{+i\lambda(\varphi-\varphi')}\right)
\delta^{\lambda}_{-\lambda'}
-m\lambda\left(k'_{\perp}e^{+i\lambda \varphi'}+
k_{\perp} e^{+i\lambda\varphi}\right)
\delta^{\lambda}_{\lambda'}\right]
\]}
\\\hline
$\gamma_{\perp}^1$ & 
\parbox{12.5cm}{
\[
\left(\frac{k'_{\perp}}{k^{'+}}e^{-i\lambda\varphi'}+
\frac{k_{\perp}}{k^+}e^{+i\lambda\varphi}
\right)\delta^{\lambda}_{-\lambda'}
-m\lambda\left(\frac{1}{k^{'+}}+\frac{1}{k^+}
\right)\delta^{\lambda}_{\lambda'}
\]}
\\\hline
$\gamma_{\perp}^2$ &
\parbox{12.5cm}{
\[ 
i\lambda\left(\frac{k'_{\perp}}{k^{'+}}e^{-i\lambda\varphi'}-
\frac{k_{\perp}}{k^+}e^{+i\lambda\varphi}
\right)\delta^{\lambda}_{-\lambda'}
-im\left(\frac{1}{k^{'+}}+\frac{1}{k^+}
\right)\delta^{\lambda}_{\lambda'}
\]}\\
\hline
\end{tabular}
\caption[Matrix elements of the Dirac spinors]
{Matrix elements of the Dirac spinors.}
\end{table}

\subsection{The helicity table}

For the calculation of matrix elements of effective interaction
in the annihilation channel 
we use the matrix elements of the Dirac spinors 
listed in Table $4$ \cite{LeBr}.  
Also the following holds
$ (\bar{v}_{\lambda'}(p)\gamma^{\alpha}u_{\lambda}(q))^{+}
 = \bar{u}_{\lambda}(q)\gamma^{\alpha}v_{\lambda'}(p)$.

\noindent
We introduce
\begin{eqnarray}
 2H^{(1)}(x,\vec{k}_{\perp};\lambda_1,\lambda_2|x',\vec{k}_{\perp}^{'};
 \lambda'_1,\lambda'_2)
 &=& \langle\gamma^{\mu}\gamma^{\nu}\rangle g_{\mu\nu}^{\perp}
  = -\langle\gamma_1^2\rangle-\langle\gamma_2^2\rangle
\nonumber\\
 2H^{(2)}(x,\vec{k}_{\perp};\lambda_1,\lambda_2|x',\vec{k}_{\perp}^{'};
 \lambda'_1,\lambda'_2)
 &=& \langle\gamma^{\mu}\gamma^{\nu}\rangle 
  \eta_{\mu}\eta_{\nu}\frac{1}{p^{+2}}
\label{eq:a16}\end{eqnarray}
where
\begin{eqnarray}
  \langle\gamma^{\mu}\gamma^{\nu}\rangle
= \frac{(\bar{v}(1-x',-\vec{k}_{\perp}^{'};\lambda'_2)\gamma^{\mu}
  u(x',\vec{k}_{\perp}^{'};\lambda'_1))}{\sqrt{x'(1-x')}}
  \frac{(\bar{u}(x,\vec{k}_{\perp};\lambda_1) \gamma^{\nu}
  v(1-x,-\vec{k}_{\perp};\lambda_2)}{\sqrt{x(1-x)}}
\label{eq:a14}\end{eqnarray}
These functions are displayed in the Table $5$.

\begin{table}[htb]
\begin{tabular}{|c||c|c|c|c|}\hline
\rule[-3mm]{0mm}{8mm}{\bf final:initial} & $(\lambda'_1,\lambda'_2)
=\uparrow\uparrow$ 
& $(\lambda'_1,\lambda'_2)=\uparrow\downarrow$ 
& $(\lambda'_1,\lambda'_2)=\downarrow\uparrow$ &
$(\lambda'_1,\lambda'_2)=\downarrow\downarrow$ \\ \hline\hline
\rule[-3mm]{0mm}{8mm}$(\lambda_1,\lambda_2)=\uparrow\uparrow$ & 
$H_1(1,2)$   
&$H_3(2,1)$ & $H^*_3(2,1)$ & $0$ \\ \hline
\rule[-3mm]{0mm}{8mm}$(\lambda_1,\lambda_2)=\uparrow\downarrow$ & 
$H_3(1,2)$ 
& $H^*_2(1,2)$ & $H_4(2,1)$ &$0$ \\ \hline
\rule[-3mm]{0mm}{8mm}$(\lambda_1,\lambda_2)=\downarrow\uparrow$& 
$H_3^*(1,2)$ & $H_4(1,2)$ & $H_2(1,2)$  & $0$\\ \hline
\rule[-3mm]{0mm}{8mm} $(\lambda_1,\lambda_2)=\downarrow\downarrow$ & $0$ 
& $0$ & $0$ & $0$  \\
\hline
\end{tabular}
\caption[General helicity table defining the effective interaction
in the annihilation channel.]
{\protect\label{GeneralHelicityTableAnnihilation}General helicity table 
defining the effective interaction in the annihilation channel.}
\end{table}
\vspace{0.5cm}

Here, the matrix elements 
$H^{(n)}_i(1,2)=H^{(n)}_i(x,\vec{k}_{\perp};x',\vec{k}_{\perp}^{'})$
are the following
\begin{eqnarray}
    H^{(1)}_1(x,\vec{k}_{\perp};x',\vec{k}_{\perp}^{'})
&=& - m^2\left(\frac{1}{x}+\frac{1}{1-x}\right)
         \left(\frac{1}{x'}+\frac{1}{1-x'}\right)
\nonumber\\
    H^{(1)}_2(x,\vec{k}_{\perp};x',\vec{k}_{\perp}^{'})
&=& - k_{\perp}k'_{\perp}\left( \frac{{\rm e}^{ i(\varphi-\varphi^{'})}}{xx'} \right)
\nonumber\\
    H^{(1)}_3(x,\vec{k}_{\perp};x',\vec{k}_{\perp}^{'})
&=& - m\lambda_1\left(\frac{1}{x}+\frac{1}{1-x}\right)
                 \frac{k_{\perp}^{'}}{1-x'}{\rm e}^{i\varphi}
\nonumber\\
    H^{(1)}_4(x,\vec{k}_{\perp};x',\vec{k}_{\perp}^{'})
&=&   k_{\perp}k'_{\perp}
           \left( \frac{{\rm e}^{i(\varphi-\varphi^{'})}}{x'(1-x)} \right)
\label{eq:a17}\end{eqnarray}

\noindent
and
\begin{eqnarray}
&& H^{(2)}_1(x,\vec{k}_{\perp};x',\vec{k}_{\perp}^{'})
=H^{(2)}_3(x,\vec{k}_{\perp};x',\vec{k}_{\perp}^{'})=0
\nonumber\\
&& H^{(2)}_2(x,\vec{k}_{\perp};x',\vec{k}_{\perp}^{'})
=H^{(2)}_4(x,\vec{k}_{\perp};x',\vec{k}_{\perp}^{'})=2
\label{eq:a15}\end{eqnarray}

\subsection{The helicity table for $|J_z|\leq 1$}

The matrix elements of the effective interaction
for $J_z\geq 0$~~ 
$F(x,k_{\perp};\lambda_1,\lambda_2|x',k'_{\perp};
\lambda'_1,\lambda'_2)=
\langle x,k_{\perp};J_z,\lambda_1,\lambda_2|\tilde{V}_{eff}|
x',k'_{\perp};J'_z,\lambda'_1,\lambda'_2\rangle$ in the annihilation channel
(the sum of the generated interaction for $J_z=+1$
and instantaneous graph for $J_z=0$)
are given in Table $6$.

\begin{table}[htb]
\begin{tabular}{|c||c|c|c|c|}\hline
\rule[-3mm]{0mm}{8mm}{\bf final:initial} & $(\lambda'_1,\lambda'_2)
=\uparrow\uparrow$ 
& $(\lambda'_1,\lambda'_2)=\uparrow\downarrow$ 
& $(\lambda'_1,\lambda'_2)=\downarrow\uparrow$ &
$(\lambda'_1,\lambda'_2)=\downarrow\downarrow$ \\ \hline\hline
\rule[-3mm]{0mm}{8mm}$(\lambda_1,\lambda_2)=\uparrow\uparrow$ & 
$F_1(1,2)$   
&$F_3(2,1)$ & $F^*_3(2,1)$ & $0$ \\ \hline
\rule[-3mm]{0mm}{8mm}$(\lambda_1,\lambda_2)=\uparrow\downarrow$ & 
$F_3(1,2)$ 
& $F^*_2(1,2)$ & $F_4(2,1)$ &$0$ \\ \hline
\rule[-3mm]{0mm}{8mm}$(\lambda_1,\lambda_2)=\downarrow\uparrow$& 
$F_3^*(1,2)$ & $F_4(1,2)$ & $F_2(1,2)$  & $0$\\ \hline
\rule[-3mm]{0mm}{8mm} $(\lambda_1,\lambda_2)=\downarrow\downarrow$ & $0$ 
& $0$ & $0$ & $0$  \\
\hline
\end{tabular}
\caption[Helicity table of the effective interaction
in the annihilation channel for $J_z\ge 0$]
{\protect\label{HelicityTableAnnihilation}Helicity table 
of the effective interaction
in the annihilation channel for $J_z\ge 0$.}
\end{table}
\vspace{0.5cm}

The function $F_i(1,2)=F_i(x,k_{\perp};x',k'_{\perp})$ are 
the following
\begin{eqnarray}
  F_1(x,k_{\perp};x',k_{\perp}^{'})
&=& \frac{\alpha}{\pi}\frac{1}{\Omega}
    \frac{m^2}{xx'(1-x)(1-x')}\delta_{|J_z|,1}
\nonumber\\
  F_2(x,k_{\perp};x',k_{\perp}^{'})
&=& \frac{\alpha}{\pi}\left(\frac{1}{\Omega}
    \frac{k_{\perp}k'_{\perp}}{xx'}\delta_{|J_z|,1}
    +2\delta_{J_z,0}\right)
\nonumber\\
  F_3(x,k_{\perp};x',k_{\perp}^{'})
&=& \frac{\alpha}{\pi}\frac{1}{\Omega}
    \lambda_1\frac{m}{x'(1-x')}
    \frac{k_{\perp}}{1-x}\delta_{|J_z|,1}
\nonumber\\
  F_4(x,k_{\perp};x',k_{\perp}^{'})
&=& \frac{\alpha}{\pi}\left(-\frac{1}{\Omega}
    \frac{k_{\perp}k'_{\perp}}{x(1-x')}\delta_{|J_z|,1}
    +2\delta_{J_z,0}\right)
\label{eq:a12a}\end{eqnarray}
where we have introduced

\noindent
(1) {\bf Exponential cut-off} 
\begin{eqnarray}
 \frac{1}{\Omega} = \frac{1}{M^2}
\,,\end{eqnarray}

\noindent
(2) {\bf Gaussian cut-off}
\begin{eqnarray}
 \frac{1}{\Omega} = \frac{M_a^2+M_b^2}{M_a^4 + M_b^4}
\,,\end{eqnarray}

\noindent
(3) {\bf Sharp cut-off}
\begin{eqnarray}
\frac{1}{\Omega} &=& \frac{\theta(M_a^2-M_b^2)}{M_a^2}
                  +  \frac{\theta(M_b^2-M_a^2)}{M_b^2}
\,.\end{eqnarray}

The table for $J_z=-1$ is obtained by inverting all helicities, i.e.
\begin{eqnarray}
&& F(J_z=+1;\lambda_1,\lambda_2)=-\lambda_1 F(J_z=-1;-\lambda_1,-\lambda_2)
\,,\label{eq:a18}\end{eqnarray}

The matrix elements of the effective interaction 
in the annihilation channel are nonzero only for $|J_z|\leq 1$
due to the restriction on the angular momentum of the photon.


\newpage
\begin{thebibliography}{99}
\bibitem {bpp97} S.J.~Brodsky, H.C.~Pauli, and S.S.~Pinsky, 
   {\it Quantum chromodynamics 
      and other field theories on the light cone},
      Physics Reports {\bf 301}, 299 (1998).
\bibitem{BrPe} 
      M.~Brisudova, and R.J.~Perry, 
      hep-ph/9511443,
      Phys.Rev. {\bf D54}, (1996) 1831;  
      %
      M.~Brisudova, R.J.~Perry, K.G.~Wilson, 
      hep-ph/9607280,
      Phys.Rev.Lett.{\bf 78}, (1997) 1227.
\bibitem{BrPe2} 
      M.~Brisudova, and R.J.~Perry, hep-ph/9605363,
      Phys.Rev. {\bf D54}, (1996) 6453. 
\bibitem{JoPeGl}
     B.~D.~Jones, R.~G.~Perry and S.~D.~Glazek, 
     Phys.Rev. {\bf D55}, 6561 (1997); hep-th/9605231; 
     B.~D.~Jones and R.~G.~Perry, hep-th/9703106.
\bibitem{Perry}
     R.J.~Perry, in {\it Proceedings of Hdrons 94}, 
     edited by V.~Herscovitz and C.~Vasconcellos
     (World Scientific, Singapore, 1995), 
     hep-th/9407056.
\bibitem{GlWi} 
       S.D. Glazek and K.G. Wilson, 
       Phys.Rev. {\bf D48}, 5863 (1993);
       S.D. Glazek and K.G. Wilson, 
       Phys.Rev. {\bf D49}, 4214 (1994).
\bibitem{weg94} 
       F.~Wegner, Ann.Physik {\bf 3},77 (1994).
\bibitem{pau96} 
  H.C.~Pauli,
      hep-th/9608035;
      hep-th/9707361; 
      hep-th/9809005,  
      MPIH-V21-1998, June 1998, to appear in Europhys.Journal (1998).
\bibitem{LeBr} 
       G.P.~Lepage, S.J.~Brodsky, 
       Phys.Rev. {\bf D22}, 2157 (1980). 
\bibitem{TrPa} 
      U.~Trittmann and H.C.~Pauli, 
      hep-th/9704215, hep-th/9705021, hep-th/9705072.
\bibitem{TrPa2} 
      Computer code is available on the World Wide Web (WWW) under
      {\bf http://pluto.mpi-hd.mpg.de/\~{}trittman/code.html}.
\bibitem{MePa} 
      U.~Merkel and H.C.~Pauli, hep-th/9608152, 
      Phys.Rev. {\bf D55}, (1997), 2486. 
\bibitem {kpw92} 
   M.~Krautg\"artner, H.C.~Pauli and F.~W\"olz, 
   Phys.Rev. {\bf D45}, (1992) 3755. 
\bibitem {KaPi}
    M. Kalu\v za and H.-J. Pirner,   
    Phys.Rev. {\bf D47}, 1620 (1993).
\bibitem {previous}
    E.L. Gubankova, H.C. Pauli, F. Wegner
    hep-th/9809143 
\bibitem{gub98} 
       E.L.~Gubankova, 
       hep-th/9801018.
\end{thebibliography}
\end{document}